\documentclass{article}
\usepackage[utf8]{inputenc}
\usepackage[british]{babel}
\usepackage{amssymb,amsmath,amsthm,amsfonts}
\usepackage{longtable}
\usepackage{placeins}
\usepackage{graphicx}
\usepackage{url}
\usepackage{comment}

\newenvironment{myenumerate}{
\begin{enumerate}
 \setlength{\itemsep}{1pt}
 \setlength{\parskip}{0pt}
 \setlength{\parsep}{0pt}}{\end{enumerate}
}

\newenvironment{myitemize}{
\begin{itemize}
 \setlength{\itemsep}{1pt}
 \setlength{\parskip}{0pt}
 \setlength{\parsep}{0pt}}{\end{itemize}
}

\theoremstyle{definition}
\newtheorem{definition}{Definition}[section]

\begin{document}

\title{\vspace{-4.2cm} Coding-theorem Like Behaviour and Emergence of the Universal Distribution from Resource-bounded Algorithmic Probability}
\author{%
Hector Zenil$^{1,2,3}$\thanks{Corresponding author: hector.zenil [at] algorithmicnaturelab [dot] org}\,
\and Liliana Badillo\,$^{4}$
\and Santiago Hern\'andez-Orozco\,$^{2,3,4}$
\and and Francisco Hern\'andez-Quiroz\,$^5$
}

\date{}

\maketitle

\thanks{%
\noindent 
$^{1}$ Department of Computer Science, University of Oxford, Oxford, U.K.\\
$^{2}$ Algorithmic Dynamics Lab, Unit of Computational Medicine, SciLifeLab, Centre for Molecular Medicine, Department of Medicine Solna, Karolinska Institute, Stockholm, Sweden.\\
$^{3}$ Algorithmic Nature Group, LABORES, Paris, France.\\
$^{4}$ Posgrado en Ciencias
e Ingenier\'ia de la Computaci\'on, Universidad Nacional Autonoma de M\'exico (UNAM)\\
$^{5}$ Departamento de Matem\'aticas, Facultad de Ciencias, Universidad Nacional Aut\'onoma de M\'exico (UNAM), Ciudad de M\'exico, M\'exico.\\}

\begin{abstract} Previously referred to as `miraculous' in the scientific literature because of its powerful properties and its wide application as optimal solution to the problem of induction/inference, (approximations to) Algorithmic Probability (AP) and the associated Universal Distribution are (or should be) of the greatest importance in science. Here we investigate the emergence, the rates of emergence and convergence, and the Coding-theorem like behaviour of AP in Turing-subuniversal models of computation. We investigate empirical distributions of computing models in the Chomsky hierarchy. We introduce measures of algorithmic probability and algorithmic complexity based upon resource-bounded computation, in contrast to previously thoroughly investigated distributions produced from the output distribution of Turing machines. This approach allows for numerical approximations to algorithmic (Kolmogorov-Chaitin) complexity-based estimations at each of the levels of a computational hierarchy. We demonstrate that all these estimations are correlated in rank and that they converge both in rank and values as a function of computational power, despite fundamental differences between computational models. In the context of natural processes that operate below the Turing universal level because of finite resources and physical degradation, the investigation of natural biases stemming from algorithmic rules may shed light on the distribution of outcomes. We show that up to 60\% of the simplicity/complexity bias in distributions produced even by the weakest of the computational models can be accounted for by Algorithmic Probability in its approximation to the Universal Distribution.\\


\end{abstract}

\section{Motivation and Significance}

Algorithmic Probability (AP) and its associated Universal Distribution (UD) predicts the way in which strings distribute when `random' computer programs are run. This algorithmic `law' thus regulates the behaviour of the output distribution of computer programs. 

The Universal Distribution is the probability distribution that establishes how the output strings from a universal computer running a random computer program distribute. Formally, the Algorithmic Probability of a string $AP(s)$ is defined by:

\begin{equation}
\label{AP}
AP_U(s) = \sum_{p:U(p) = s} 1/2^{|p|}
\end{equation}

\noindent where the sum is over all halting programs $p$ for which $U$, a prefix-free universal Turing machine, outputs the string $s$. A prefix-free universal Turing machine defines a set of valid programs such that the sum is bounded by Kraft's inequality~\cite{kraft} and not greater than 1 (it is also called a semi-probability measure because some programs will not halt, and thus the sum of the probabilities is never really 1).

An \textit{invariance theorem} establishes that the choice of reference universal Turing machine introduces a vanishing bias as a function of string size:

\begin{equation}
\label{invariance}
AP_U(s) \leq c_{U,U^\prime} AP_{U^\prime}(s)
\end{equation}

\noindent where $c_{U,U^\prime}$ is a constant that depends on $U$ and $U^{\prime}$ (think of a compiler size translating in both directions) but is independent of $s$. Hence the reference $U$ can safely be dropped in the long term. Yet this invariance theorem tells us nothing about the rate of convergence, thus making these numerical experiments more relevant and necessary.

The Algorithmic Probability and the Universal Distribution represent the theoretically optimal response to the challenge of induction and inference, according to R. Solomonoff, one of the proponents of algorithmic information theory~\cite{solo,solo2,solo3}.

More recently, at a panel discussion at the World Science Festival in New York City on Dec 14, 2014, Marvin Minsky, one of the founding fathers of AI, said (own transcription): 

\begin{quotation}
It seems to me that the most important discovery since G\"odel was the discovery by Chaitin, Solomonoff and Kolmogorov of the concept called Algorithmic Probability, which is a fundamental new theory of how to make predictions given a collection of experiences, and this is a beautiful theory, everybody should learn it, but it's got one problem, that is, that you cannot actually calculate what this theory predicts because it is too hard, it requires an infinite amount of work. However, it should be possible to make practical approximations to the Chaitin, Kolmogorov, Solomonoff theory that would make better predictions than anything we have today. Everybody should learn all about that and spend the rest of their lives working on it.
\end{quotation}

The Universal Distribution has also been characterized as \textit{miraculous}, because of its performance in inference and prediction~\cite{kirchner}. However,  the calculation of both $AP(s)$ is not computable. This has meant that for decades after its discovery, few attempts have been made to apply Algorithmic Probability to problems in general science. However, to be more precise $AP(s)$ is upper semi-computable, which means that it can be approximated from below. 
AP is of fundamental interest to science, as it addresses the most pressing challenges in the areas of complexity, inference and causality, attempting to keep pushing towards better methods for approximating algorithmic probability. A recent new framework and a pipeline of relevant numerical methods have been advanced and have proven successful in many areas, ranging from cognition to graph complexity~\cite{ploscompbio,kempe,gauvrit2,physicaA,seminars}.

There are many properties of $m$ that make it optimal~\cite{solo,solo2,solo3,kirchner}. For example, the same Universal Distribution will work for any problem within a convergent error; it can deal with missing and multidimensional data; the data do not need to be stationary or ergodic; there is no under-fitting or over-fitting because the method is parameter-free and thus the data need not be divided into training and test sets; it is the gold standard for a Bayesian approach in the sense that it updates the distribution in the most efficient and accurate way possible with no assumptions.

Several interesting extensions of resource-bounded Universal Search approaches have been introduced in order to make algorithmic probability more useful in practice~\cite{schmid1,schmid2,godelmachines,AIXI}. Some of these provide some theoretical bounds~\cite{antunes}. Certain of these approaches have explored the effect of relaxing some of the conditions (e.g. universality) on which Levin's Universal Search is fundamentally based~\cite{wiering} or have introduced domain-specific versions (and thus versions of conditional AP). Here we explore the behaviour of explicitly weaker models of computation-- of increasing computational power-- in order to investigate asymptotic behaviour and the emergence of the Universal Distribution, and the properties of both the different models with which to approximate it and the actual empirical distributions that such models produce. 

The so-called Universal Search~\cite{levin} is based on dovetailing all possible programs and their runtimes such that the fraction of time allocated
to program $p$ is $1/2^{|p|}$, where $|p|$ is the size of the program (in number of bits). Despite the algorithm's simplicity and remarkable theoretical properties, a potentially huge constant slowdown factor has kept it from being much used in practice. Some of the approaches to speeding it up have included the introduction of bias and making the search domain specific, which has at the same time limited the power of Algorithmic Probability. 

There are practical applications of AP that make it very relevant. If one could translate some of the power of Algorithmic Probability to decidable models (thus below  Type-0 in the Chomsky hierarchy) without having to deal with the uncomputability of algorithmic complexity and algorithmic probability, it would be effectively possible to trade computing power for predictive power. While trade-offs must exist for this to be possible (full predictability and uncomputability are incompatible), the question of finding a threshold for the coding-theorem to apply would be key to transferring some power through relaxing the computational power of an algorithm. If a smooth trade-off is found before the undecidability border of Turing completeness, it would mean that the partial advantages of Algorithmic Information Theory can be found and partially recovered from simpler models of computation in exchange for accuracy. Such simpler models of computation may model physical processes that are computationally rich but are subject to noise or are bounded by resources. More real-world approaches may then lead to applications such as in the reduction of conformational distributions of protein folding~\cite{kamal1,kamal2} in a framework that may favour or rule out certain paths, thereby helping predict the most likely (algorithmic) final configuration. If the chemical and thermodynamic laws that drive these processes are considered algorithmic in any way, even under random interactions, e.g. molecular Brownian motion, the Universal Distribution may offer insights that may help us quantify the most likely regions if these laws in any sense constitute forms of computation below or at the Turing level that we explore here. This will appear more plausible if one considers the probabilistic bias affecting convergence~\cite{schaper}, bearing in mind that 
we have demonstrated that biological evolution operating in algorithmic space
can better explain some phenomenology related to natural selection~\cite{evocomplex}.

\subsection{Uncomputability in complexity}

Here we explore the middle ground at the boundary and study the interplay between \textit{computable} and \textit{non-computable} measures of algorithmic probability connected to algorithmic complexity. Indeed, a deep connection between the \textit{algorithmic complexity} (or Kolmogorov-Chaitin complexity) of an object $s$ and $AP$ of $s$ was found and formalized by way of the \textit{algorithmic Coding theorem}. The theorem establishes that the probability of $s$ being produced by a random algorithm is inversely proportional to its algorithmic complexity (up to a constant)~\cite{levin1974}:

\begin{equation}
\label{codingtheorem}
- \log AP(s) = K(s) + O(1)
\end{equation}
\label{m}

Levin proved that the output distribution established by Algorithmic Probability dominates (up to multiplicative scaling) any other distribution produced by algorithmic means as long as the executor is a universal machine, hence giving the distribution its `universal' character (and  its name: `Universal Distribution').

This so-called Universal Distribution is a signature of Turing-completeness. However, all processes that model or regulate natural phenomena may not necessarily be Turing universal. For example, some models of self-assembly may not be powerful enough to reach Turing-completeness, yet they display similar output distributions to those predicted by the Universal Distribution by way of the algorithmic Coding theorem, with simplicity highly favoured by frequency of production. Noise is another source of power degradation that may preempt universality and therefore the scope and application of algorithmic probability. However, if some subuniversal systems approach the coding-theorem behaviour, these give us great prediction capabilities, and less powerful but computable algorithmic complexity measures. Here we ask whether such distributions can be partially or totally explained by importing the relation established by the coding theorem, and under what conditions non-universal systems can display algorithmic coding-theorem like behaviour.

We produce empirical distributions of systems at each of the computing levels of the Chomsky hierarchy, starting from transducers (Type-3) as defined in~\cite{calude}, Context-free grammars (Type-2) as defined in~\cite{wharton}, linear-bounded non-deterministic Turing machines (Type-1) as approximations to bounded Kolmogorov-Chaitin complexity and a universal procedure from an enumeration of Turing machines (Type-0) as defined in~\cite{zenild4,solerzenil}. We report the results of the experiments and comparisons, showing the gradual coding-theorem-like behaviour at the boundary between decidable and undecidable systems.



\section{Methods}

We will denote by $TM(n,k)$ or simply $(n,k)$ the set of all strings produced by all the Turing machines with $n$ states and $k$ symbols.

\subsection{The Chomsky Hierarchy}

The Chomsky hierarchy is a strict containment hierarchy of classes of formal grammars equivalent to different computational models of increasing computing power. At each of the 4 levels, grammars and automata compute a larger set of possible languages and strings. The four levels, from weaker to stronger computational power, are:

\begin{myenumerate}
\item[Type-3] The most restricted grammars generating the regular languages. They consist of rules with single non-terminal symbols on the left-hand side and strings of terminal symbols followed by at most one non-terminal symbol on the right-hand side. These types of rule are referred to as right linear (but a symmetrical left linear definition works as well). This level is studied by way of finite-state transducers (FST), a generalization of finite-state automata (FSA) that produce an output at every step, generating a relation between input strings and output strings. Though apparently more general, FST-recognized languages are the same as FSA-accepted sets. Hence both can represent this level. We use an enumeration of transducers introduced in~\cite{calude} where they also proved an \textit{invariance theorem}, thus demonstrating that the enumeration choice is invariant (up to a constant).

\item[Type-2] Grammars that generate the context-free languages. These kinds of grammars are extensively used in linguistics. The languages generated by CFG grammars are exactly the languages that can be recognized by a non-deterministic pushdown automaton. We denote this level by CFG. We generated production rules for 40\,000 grammars according to a sound scheme introduced in~\cite{wharton}.

\item[Type-1] Grammars that generate the context-sensitive languages. The languages described by these grammars are all languages that can be recognized by a linear-bounded automaton (LBA), a Turing machine whose tape's length is bounded by a constant times the length of the input. An AP-based variation is introduced here, and we denote it by LBA/AP.

\item[Type-0] The least restricted grammar. Generates the languages that can be recognized by Turing machines, also called recursively enumerable languages. This is the level at which Turing-universality is achieved or required. We used previously generated distributions produced and reported in~\cite{zenild4,solerzenil}
\end{myenumerate}

We also explore the consequences of relaxing the halting configuration (e.g. halting state) in models of universal computation (Type-0) when it comes to comparing their output distributions.

\subsection{Finite-state complexity}
 
Formal language theory and algorithmic complexity had traditionally been at odds as regards the number of states, or the number of transitions in a minimal finite automaton accepting a regular language. In~\cite{campeanu} a connection was established by extending the notions of Blum static complexity and encoded function space. The main reason for this lack of connection was that languages are sets of strings, rather than strings used for measures of algorithmic complexity, and a meaningful definition of the complexity of a language was lacking, as well as a definition of Finite-state algorithmic complexity. However, \cite{calude} offered a version of algorithmic complexity by replacing Turing machines with finite transducers. The complexity induced is called Finite-state complexity (FSA). Despite the fact that the Universality Theorem (true for Turing machines) is false for finite transducers, rather surprisingly, the \textit{invariance theorem} holds true for finite-state complexity and, in contrast with descriptional complexities (plain and prefix-free), finite-state complexity is computable.

Defined in~\cite{calude} and analogous to the core concept (Kolmogorov-Chaitin complexity) of Algorithmic Information Theory (AIT)-- based on finite transducers instead of Turing machines-- finite-state complexity is computable, and there is no a priori upper bound for the number of states used for minimal descriptions of arbitrary strings.

Consider a transducer $T$ with the finite set of states $Q= \{ 1, \dots ,n \}$. Then the transition function $\Delta$ of $T$ is encoded by a binary string $\sigma$ (see \cite{calude} for details). The transducer $T$ which is encoded by $\sigma$ is called $T_{\sigma}^{S_{0}}$, where $S_{0}$ is the set of all strings in the form of $\sigma$.

In~\cite{calude} it was shown that the set of all transducers could be enumerated by a regular language and that there existed a hierarchy for more general computable encodings. For this experiment we fix $S=S_{0}$.

As in traditional AIT, where Turing machines are used to describe binary strings, transducers describe strings in the following way: we say that a pair $(T_{\sigma}^{S},p)$, $\sigma \in S$, $p \in \mathbb{B}^{*}$, where $\mathbb{B}^{*}$ is the set of all finite binary strings, is a description of the string $s \in\mathbb{B}^{*}$ if and only if $T_{\sigma}^{S}(p)=s$. The size of the description $(T_{\sigma}^{S},p)$ is defined in the following way

$$|| (T_{\sigma}^{S},p) || = |\sigma| + |p|.$$

\theoremstyle{definition}
\begin{definition}{(\cite{calude})}
The finite-state complexity of $s \in \mathbb{B}^{*}$ (that we will identify as FSA in the results) with respect to encoding $S$ is defined by 

$$C_{S}(s)= \min_{\sigma \in S, p \in \mathbb{B}^{*}} \{ || (T_{\sigma}^{S},p) || : T_{\sigma}^{S}(p)=s \}.$$
 
\end{definition}

An important feature of traditional AIT is the \textit{invariance theorem}, which states that complexity is optimal up to an additive constant and relies on the existence of a Universal Turing machine (the additive constant is in fact its size). In contrast with AIT, due to the non-existence of a ``Universal transducer'', finite-state complexity includes the size of the transducer as part of the encoding length. Nevertheless, the \textit{invariance theorem} holds true for finite-state complexity. An interesting consequence of the invariance theorem for finite-state complexity is the existence of an upper bound $C_{S}(s) \leq |s|+8$ for all $s \in \mathbb{B}^{*}$, where $8$ is the length of the string $\sigma$ which encodes the identity transducer. Hence $C_{S}$ is computable. If $S_{n_{0}}$ and $S_{n_{1}}$ are encodings then $C_{S_{n_{0}}}=f(C_{S_{n_{1}}})$ for computable function $f$~\cite{calude}. 

An alternative definition of finite-state complexity based on Algorithmic Probability is as follows:

\theoremstyle{definition}
\begin{definition}{(\cite{calude})}
The finite-state complexity (denoted by FSA/AP in the results) of $s \in \mathbb{B}^{*}$ with respect to encoding $S$ is defined by 

$$C_{S}(s)= - \log \Sigma_{\sigma \in S, p \in\mathbb{B}^{*}} \{ || (T_{\sigma}^{S},p) || : T_{\sigma}^{S}(p)=s \}.$$

That is, the number of times that a string is accepted by a transducer (in this case, reported in the results, for encodings of size 8 to 22).
\end{definition}

\subsubsection{Building a finite-state empirical distribution}

We now define the construction of an empirical distribution using finite-state complexity. We introduce our alternative definition of algorithmic probability using transducers.

\theoremstyle{definition}
\begin{definition}{(Finite-state Algorithmic Probability)}
Let $S$ be the set of encodings of all transducers by a binary string in the form of $\sigma$. We then define the algorithmic probability of a string $s \in \mathbb{B}^{*}$ as follows

$$AP_{S}(s)=\sum_{(\sigma,p):T_{\sigma}^{S}=s} 2^{-(|\sigma| + |p|)}.$$
 
\end{definition}

For any string $s$, $P_{S}(s)$ is the algorithmic probability of $s$, computed for the set of encodings $S$. In the construction of the empirical distribution for finite-state complexity, we consider the set of strings $\tau \in \mathbb{B}^{*}$ such that $\tau = \sigma^\frown p $ (this is the concatenation of the binary strings $\sigma$ and $p$), $\sigma \in S$, $p \in \mathbb{B}^{*}$ and $8 \leq |\tau| \leq 22$. Hence $|\tau|=|| (T_{\sigma}^{S},p) ||$. Following~\cite{2} we define the empirical distribution function $D_{S}(n)$ (i.e. the probability distribution) as

\theoremstyle{definition}
\begin{definition}{(Finite-state Distribution, FSA)}
$$D_{S}(n)=\frac{| \{ (\sigma,p): \sigma \in S, p \in,1 \}^{*}, s \in,1 \}^{*}, T_{\sigma}^{S}(p)=s \} |}{|\{ (\sigma,p): \sigma \in S,  p \in \mathbb{B}^{*}, || (T_{\sigma}^{S},p) ||=n \} |}.$$
\end{definition}

In other words, $D_{S}(n)$ considers all strings $\tau \in \mathbb{B}^*$ of length $n$ and determines whether $\tau = \sigma^\frown p$ such that $\sigma \in S$. Then $D_{S}(n)$ computes $T_{\sigma}^{S}(p)$ and counts the number of times that we have $T_{\sigma}^{S}(p)=s$ for every string $s$ described by $(\sigma,p)$ such that $|| (T_{\sigma}^{S},p) ||=n$ \footnote{Since $S$ is in fact regular we could indeed use an enumeration of $S$, but in this instance we analyze all binary strings of length $n$.}.  

We note that in the encoding $\sigma \in S$, a string $\nu \in \mathbb{B}^{*}$ occurring as the output of a transition in $T_{\sigma}^{S}$ contributes $2\cdot|\nu|$ to the size $|| (T_{\sigma}^{S},p) ||$ of a description of a string $s$. The decision to consider strings $\tau$ such that $8 \leq |\tau|$ was made based on the fact that in the encoding of the smallest transducer i.e. the one with the transition function,

\begin{equation} \label{transd:1}
\Delta(1,0)=\Delta(1,1)=\Delta(1,\varepsilon) 
\end{equation}

\noindent where $\varepsilon$ is the empty string, the string $\nu$ (which occurs as the output of transitions in $T_{\sigma}^{S}$) has length $|\nu|=4$ and so contributes $2 \cdot |\nu|=8$ to the size $|| (T_{\sigma}^{S},p) ||$ of the description of a string $s$.

\subsection{Context-free grammars}

In~\cite{wharton} Wharton describes an algorithm for a general purpose grammar enumerator which is adaptable to different classes of grammars (i.\,e. regular, context-free, etc). We implemented this algorithm in the Wolfram Language with the purpose of enumerating context-free grammars over the terminal vocabulary $\{ 0,1 \}$, which are in Chomsky Normal Form. Before describing the implementation of the algorithm we define the following terminology:

\begin{myitemize}
\item A grammar $G$ is a 4-tuple $(N,T,P,X)$.
\item $N$ is the non-terminal vocabulary.
\item $T$ is the terminal vocabulary.
\item $P$ is the set of productions.
\item $X$ is the start symbol.
\item $V=N \cup T$ is the vocabulary of $G$.
\item For any grammar $G$, $n,t$ and $p$ denote the cardinalities of $N,T$ and $P$ respectively.

\end{myitemize} 

First, we define the structure of a grammar $G$. Let $G$ be any grammar. Suppose we are given the non-terminal vocabulary $N$ with an arbitrary ordering such that the first non-terminal is the start symbol $X$. The grammar $G$ has a structure $S$ which consists of a list of $n$ integers. Each integer $S_{i}$ from $S$ is the number of productions having the $ith$ non-terminal on the left-hand side (according to the ordering of $N$). Hence the cardinality $p$ of the set of productions $P$ satisfies $\Sigma_{i=0}^{n} S_i=p$.

Now, let $\Gamma$ be a class of grammars over a terminal vocabulary $T$. By $\Gamma_c$ we denote the grammars in $\Gamma$ with complexity $c$. We then enumerate $\Gamma$ by increasing the complexity $c$. To every complexity class $\Gamma_c$ corresponds a set of structure classes which is determined by $n$ and $p$. Therefore a complexity class $\Gamma_c$ is enumerated by enumerating each of its structure classes $\Gamma_{c,S}$ (i.\,e. every structure $S$ that constitutes $\Gamma_c$). In addition, we need to define an ordered sequence $R$ which consists of all possible right-hand sides for the production rules. The sequence $R$ is ordered lexicographically (first terminals, then non- terminals) and is defined according to the class of grammars we want to enumerate. For example, suppose we are interested in enumerating the class of Chomsky Normal Form grammars over the terminal vocabulary $T=\{0,1\}$ and the non-terminal vocabulary $N=\{X,Y\}$, we then set $R=\{0,1,XX,XY,YX,YY \}$.

\subsubsection{Implementation}
\label{grammarimplementation}

Given a complexity $c$, the algorithm described below (that we implemented in the Wolfram Language running on Mathematica) enumerates all the grammars according to~\cite{wharton} in a structure class $\Gamma_{c,S}$.

\begin{myenumerate}
\item The complexity measure $c$ is provided by the following pairing function in $p$ and $n$:
$$c=((p+n-1)(p+n-2)/2)+n.$$ In other words, given $c$, we apply the inverse of the above function in order to obtain the values of $p$ and $n$. This function is implemented by the function pairingInverse[$c$].
\item The set of non-terminals $N$ is generated by the function generateSetN[$\{n,p\}$]. 
\item The ordered sequence $R$ is generated using the set of non-terminals $N$ by the function generateSetR[$N$].
\item The different structure classes $\Gamma_{c,S}$ that correspond to complexity $c$ are generated by the function generateStructureClasses[$\{n,p\}$].
\item All the possible grammars with the structure classes defined in the previous step are then generated. Each grammar has an associated matrix $A$. This is performed by function generateStructureMatricesA[$S$, Length[$R$]]. 
\item The sequence $R$ is used to generate the rules of the grammars by the function generateGrammars[$matricesA$, $R$]. 
\end{myenumerate} 

\subsubsection{The CYK algorithm}
\label{cyk}

A procedure to decide if a string $s$ is generated by a grammar $G$ in polynomial time was implemented according to the Cocke-Younger-Kasami (CYK) algorithm. The CYK is an efficient worst-case parsing algorithm that operates on grammars in Chomsky normal form (CNF) in $O(n^3\cdot|G|)$, where $n$ is the length of the parsed string and $|G|$ is the size of the CNF grammar $G$. The algorithm considers every possible substring of the input string and decides $s \in L(G)$, where $L(G)$ is the language generated by $G$. The implementation was adapted from~\cite{mondragon}.

\subsection{CFG Algorithmic Probability}

We can now define the Algorithmic Probability of a string according to CFG as follows:

\theoremstyle{definition}
\begin{definition}{(Context-free Algorithmic Probability, CFG)}

We define a CFG empirical distribution in terms of the Universal Distribution as follows:

$$D_{s}(c)=\frac{|\{G:s \in L(G)\}|}{|\{G\}|}$$

\end{definition}

\noindent where $c=((p+n-1)(p+n-2)/2)+n$ as defined in~\ref{grammarimplementation}, $L(G)$ is the language generated by $G$ and $|\{G\}|$ denotes the cardinality of the sample set of the grammars considered. For the results here reported $|\{G\}|= 40\,000$, where all the grammars have a complexity at most $c$ according to a structure class $\Gamma_{c,S}$~\cite{wharton}. 'Complexity' here simply means a measure of the grammar size defined as a product of the grammar parameters.

\subsection{Linear-bounded complexity}

In~\cite{antunes} it is shown that the time-bounded Kolmogorov distribution is universal (in the sense of convergence), and they cite the question of an analogue to the algorithmic Coding theorem as an open problem possibly to be tackled by exploiting the universality finding. On the other hand, in~\cite{bienvenu,holzl} it has been shown that the time-bounded algorithmic complexity (being computable) is a Solovay function. These functions are an upper bound of algorithmic complexity (prefix-free version) and they give the same value for almost all strings.

In~\cite{zenild4,solerzenil} we described a numerical approach to the problem of approximating the algorithmic complexity for short strings. This approach does an exhaustive execution of all deterministic 2-symbol Turing Machines, constructs an output frequency distribution, and then applies the Coding Theorem to approximate the algorithmic complexity of the strings produced.

For this experiment we follow the same approach using a less powerful model of computation, namely, linear-bounded automata (\textit{LBA}).
A \textit{LBA} is basically a single tape Turing Machine which never leaves those cells on which the input was placed~\cite{2}. It is well known that the class of languages accepted by \textit{LBA} is in fact the class of context-sensitive languages~\cite{2}.

We use the same Turing Machine formalism as \cite{zenild4,solerzenil}, which is that of the Busy Beaver introduced by Rado~\cite{rado}, and we use the known values of the Busy Beaver functions.

\subsection{Time complexity}

\textit{\(PTime\)} or \textit{\(P\)} is the class of Turing Machines that produce an output in polynomial time with respect to the size of its input. When considering empty input strings (in machines with the same number of states), it is easy to see that this class is contained by the class defined by Linear Bounded Automata (LBA): if the number of the transition is bounded by a linear function, so are the number of cells it can visit. But it is important to note that LBA are not time restricted and can use non-deterministic transitions. Whether \(PTime\) is a subset of $NSPACE(O(n))$ (LBA) in other cases is an open question. Now, given that Turing Machines can decide context-free grammars in polynomial time (Subsection~\ref{cyk}), \(PTime\) is higher in the hierarchy than Type-2 languages.

Within the context of this article, we will represent this class with the set of Turing Machines with 4 states and 2 symbols with no inputs whose execution time is upper-bounded by a fixed constant. We will cap our execution time by 27, 54, 81 and 107 for a total of 44\,079\,842\,304 Turing machines, where 107 is the Busy Beaver value of the set.

\subsection{The Chomsky hierarchy bounding execution
time}\label{capturing-chomskys-heriachy-by-bounding-the-execution-time}

The definition of bounded algorithmic complexity is a variation of the unbounded version, as follows:

\theoremstyle{definition}
\begin{definition}{(Linear-bounded Algorithmic Complexity, CFG)}
$$C^t(s) = \min\{|p|: \textit{$U(p)=s$ and $U(p)$ runs in at most $t$ steps}\}$$
\end{definition}

Being bounded by polynomial-sized tapes, the Turing Machines that decide context-sensitive grammars (type-1) can be captured in exponential time by deterministic Turing Machines.

Exactly where each class of the Chomsky hierarchy is with respect to the time-based computational complexity classification is related to seminal open problems. For instance, a set equality between the languages recognized by linear-bounded automata and the ones recognized in exponential time would solve the \(PSPACE\not=EXPTIME\)
question. Nevertheless, varying the allowed computation time for the CTM algorithm allows us to capture approximations to the descriptive complexity of an object with fewer computing resources, in a similar fashion to considering each member of the Chomsky hierarchy.

\subsection{Non-halting models}

A first comparison between halting Turing machines and non-halting cellular automata was conducted in~\cite{algonature}, where it was reported a strong correlation but also some differences, here we further investigate and compare the distributions of halting and non-halting models in connection to the emergence of algorithmic probability and the coding-theorem like behaviour in variations of Turing-complete models.

Here, we also consider models with no halting configuration, such as cellular automata (nonH-CA) and Turing machines (nonH-TM) with no halting state as defined in~\cite{wolfram}, in order to assess whether or not they converge to the Universal Distribution defined over machines with a halting condition. For cellular automata we exhaustively ran all the 256 Elementary Cellular Automata~\cite{wolfram} (i.e. closest neighbour and centre cell are taken into consideration) and all 65\,536 so-called General Cellular Automata~\cite{wolfram} (that is, with 2 neighbours to one side and one to the other, plus the centre cell). For Turing machines, we ran all 4096 (2,2) Turing machines with no halting state, and a sample of 65\,536 (same number as CA) Turing machines in (3,2), also with no halting state.

\subsection{Consolidating the empirical distributions}

In order to perform the comparisons among the distributions in each of the Chomsky hierarchy levels, it is necessary to consolidate cases of bias imposed by arbitrary choices of parameter models (e.g. starting from a tape with 0 or 1). This is because, for example, the string 0000 should occur exactly the same number of times as 1111 does and so on, because 0000 and 1111 should have the same algorithmic complexity. If $s$ is the string and $f(s)$ the frequency of production, we thus consolidate the algorithmic probability of $s$ denoted by $AP(s)$ as follows:

$$AP(s)=\frac{f(s)+f(r(s))+f(c(s))+f(r(c(s))}{|g|}$$

\noindent where $|g|$, by Burnside's counting theorem, is the number of non-invariant transformations of $s$ and $r(s)$ is the reversion of $s$, e.g. 0001 becomes 1000, and $c(s)$ is the negation of $s$, e.g. 0001 becomes 1110, for all empirical distributions for FSA, CFG, LBA and TM. It is worth noting that $c$ and $r$ do not increase the algorithmic complexity of $s$ except by a very small constant, and thus there is no reason to expect either the model or the complexity of the strings to have been produced or selected 
to have different algorithmic complexities. More details on the counting method are given in~\cite{zenild4}.

Notice that this consolidation is not artificially imposed. It would be natural to run each computational model for each possible parameter from the model's parameter space (e.g. running all TMs once on a `blank symbol' as 0 and then again for the blank as 1) and the distribution would balance itself accordingly. So the application of the consolidating method is only to avoid unnecessary computation.

\section{Results}

Here we report the shape and rank of strings in each of the empirical distributions produced by the models of computation, by increasing computational power according to the Chomsky hierarchy and by time computational hierarchy, including comparisons to the performance of classical Shannon entropy (over the bit strings, assuming a uniform distribution) and of lossless compression (using the Compress algorithm), as well as the behaviour of 2 non-halting models of computation. Table~\ref{no_strings} reports the number of strings produced by each model. 

\begin{table}[!htb]
\centering
\begin{tabular}{c|c|c}
  \hline
Chomsky  & Computational & No. \\
Type & Model & Strings\\
  \hline
3 & FSA(8-22) & 294 \\
3 & FSA/AP(8-22) & 1038\\
2 & CFG(40K) & 496\\
(2,0) & LBA(27) & 847\\
(2,0) & LBA(54) & 1225\\
(2,0) & LBA(81) & 1286\\
0 & LBA 107 = TM(4,2) & 1302\\
0 & TM(5,2) & 8190\\
   \hline
\end{tabular}
\caption{Strings of at most 12 bits produced at each level from a total of 8190 possible strings (all generated by $(5,2)$ except 2~\cite{solerzenil}). LBA were simulated by halting time cutoffs, which is strictly less powerful than the TM model (Type-0) but strictly greater than CFG (Type-2). Each LBA is followed by its runtime cutoff. The cutoff value of 107 steps is the Busy Beaver for (4,2), and given that LBA are emulated by TM in (4,2) by letting them run up to 107, this means that they have exhausted the full TM space (4,2) and thus are equivalent. The empirical distribution for CFG was obtained by producing 40\,000 grammars and checking whether each of the 1302 strings occurring in LBA 107 could be generated by any of these grammars. A Chomsky Type (2,0) means that the model is strictly between Type-0 and Type-2, thus representing Type-1.}\label{no_strings}
\end{table}

\begin{figure}[htp]
\centering
\textbf{A}\hspace{5.5cm}\textbf{B}\\
\medskip

\includegraphics[width=5.5cm]{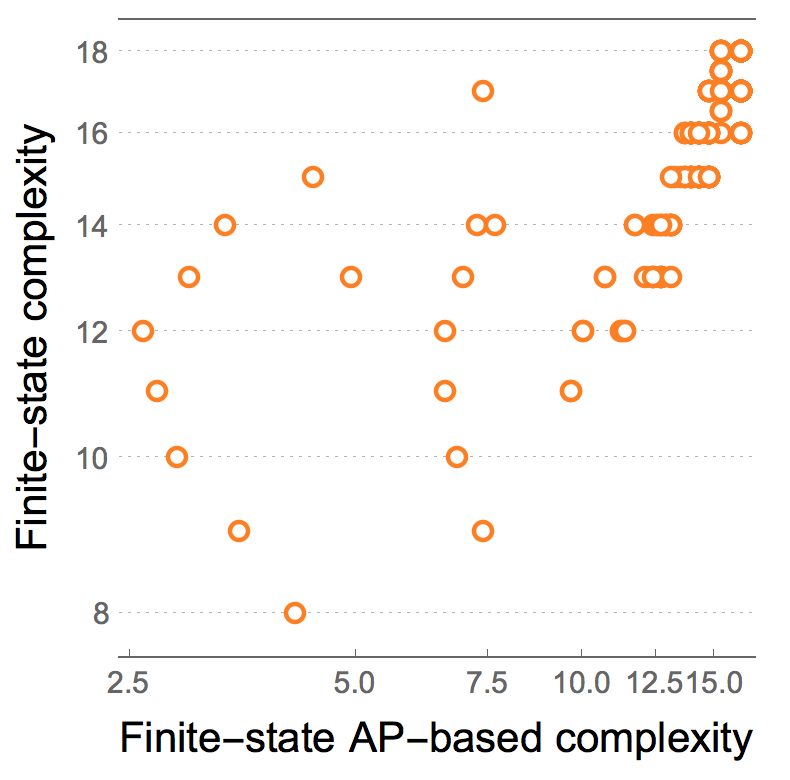}\hspace{.6cm}\includegraphics[width=5.7cm]{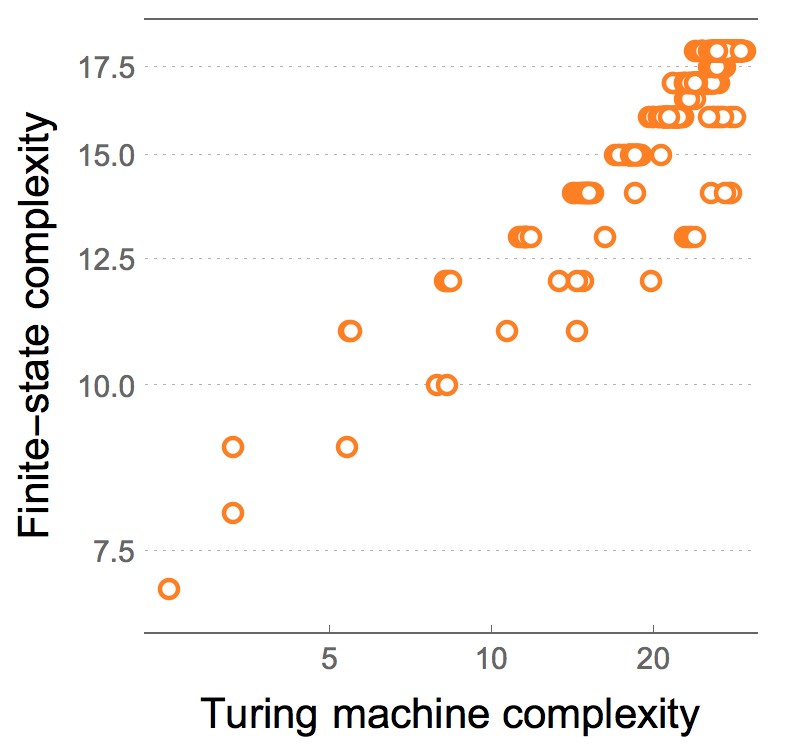}\\
\bigskip
\bigskip

\textbf{C}\hspace{2cm}\\
\medskip

\includegraphics[width=5.2cm]{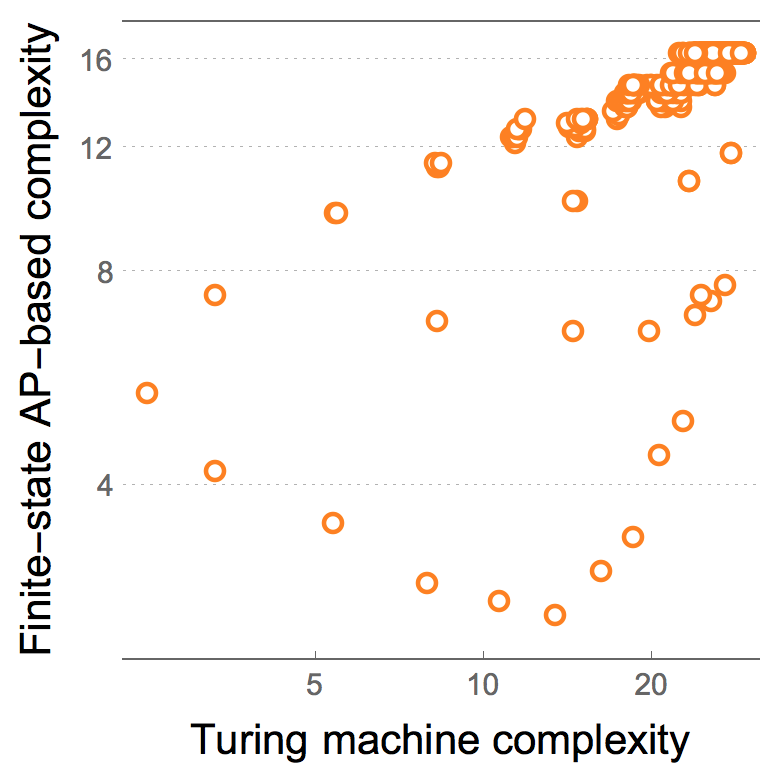}
 \caption{\label{cora}Correlation plots (A) between the 2 models of Algorithmic Probability for FSA using the same enumeration (B) between FSA (both plain and FSA/AP-based) and (C) both FSA and FSA/AP separately against Turing machines.}
\end{figure}

\begin{figure}[htp]
\centering
\textbf{A}\hspace{5.5cm}\textbf{B}\\
\medskip

\includegraphics[width=4.5cm]{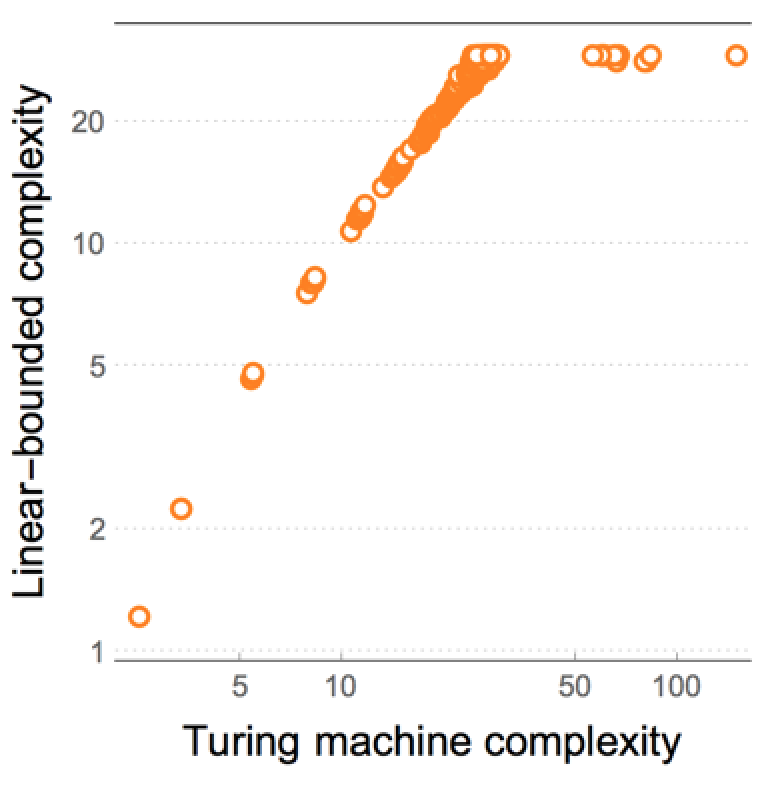}\hspace{.6cm}\includegraphics[width=6.5cm]{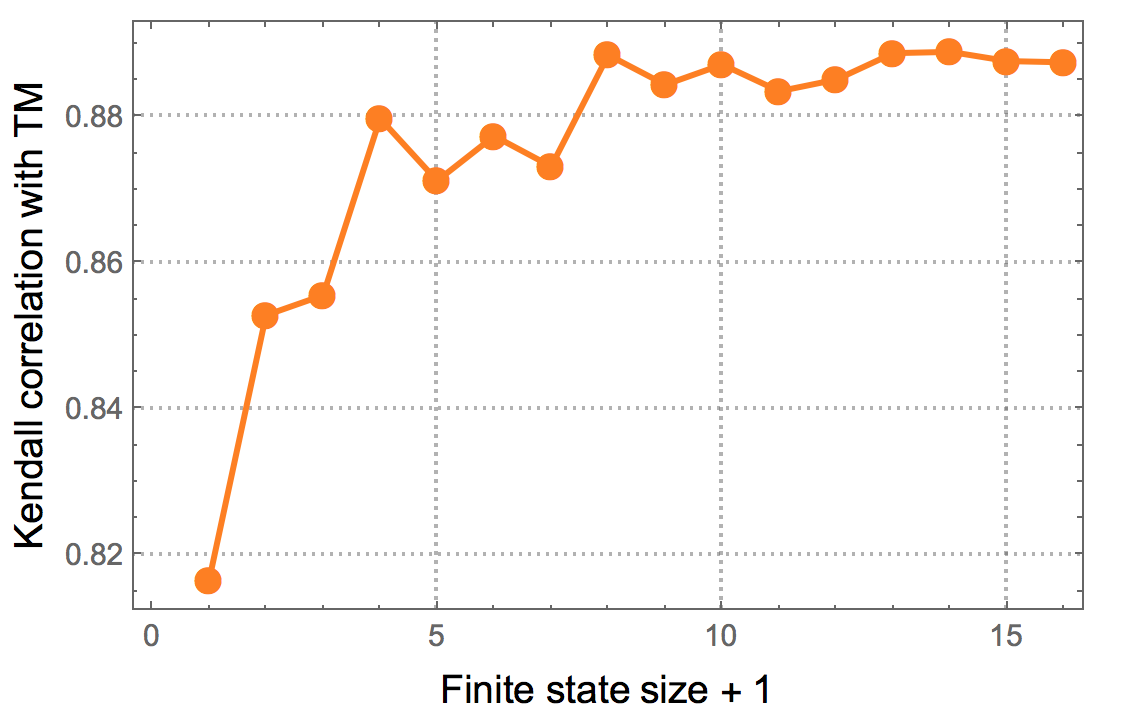}\\
\bigskip
\bigskip

\textbf{C}\hspace{5.3cm}\textbf{D}\\
\medskip

\includegraphics[width=5cm]{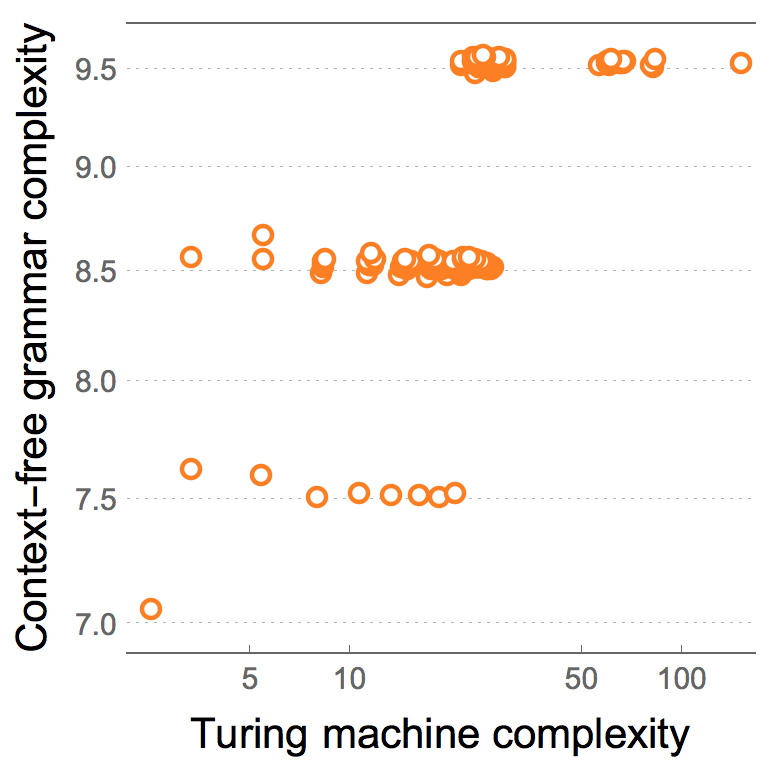}\hspace{1.9cm}\includegraphics[width=5cm]{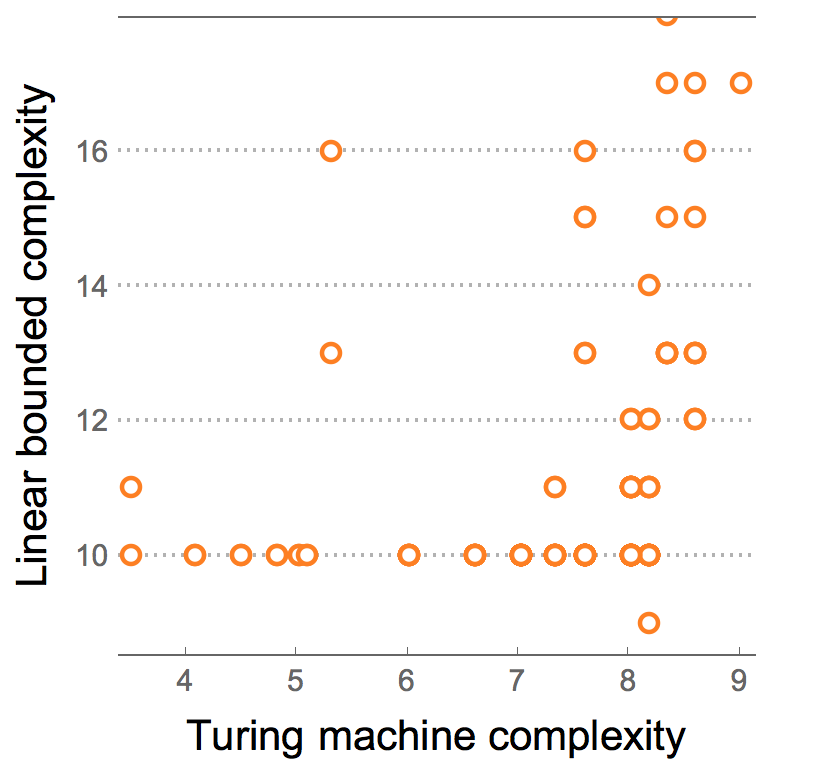}
 \caption{\label{cor}Correlation plots of the four Chomsky hierarchy types. (A) FSA v TM. (B) how FSA approximates TM before converging to a Kendall ranking correlation value below 0.9 (with LBA able to reach 1, see Fig.~\ref{timedist}). (C) CFG v TM and (D) LBA v TM.}
\end{figure}

\subsection{Finite-state complexity}

The experiment consists of a thorough analysis of all strings $\tau \in,1 \}^{*}$ that satisfy $8 \leq |\tau| \leq 22$. If the string $\tau$ satisfies $\tau = \sigma^\frown p$ (for some $\sigma \in S$) then we compute $T_{\sigma}^{S}(p)$ and generate a set of output strings. Then a frequency distribution is constructed from the set of output strings. On the other hand, we compute the finite-state complexity for strings $s$ such that $0 \leq s \leq 8$ (this is an arbitrary decision).

\subsubsection{distributions $D_{S}(n)$ produced for $8 \leq n \leq 22$}

The results given in Table~\ref{table:validtrans} indicate how many strings $\tau$ satisfy  $\tau = \sigma^\frown p$ (such that $\sigma$ encodes the transition table of some transducer $T_{\sigma}^{S}$ and $p$ is an input for it) per string length.

\begin{figure}[ht]
\centering
\includegraphics[width=9cm]{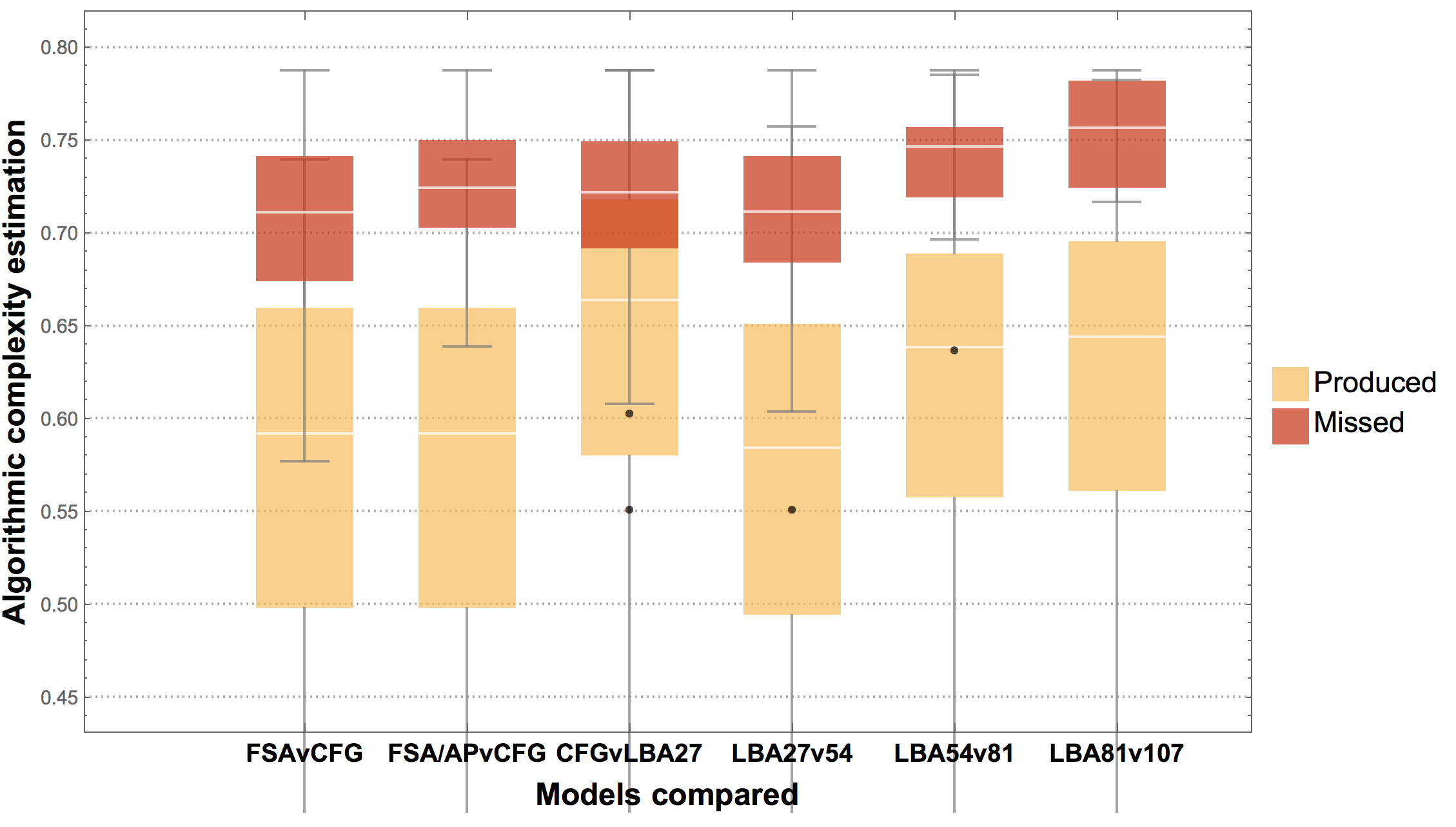} \caption{\label{missed}Each model misses the most random strings according to the next most powerful model, with the greatest discrepancy against TM(5,2). For example, FSA and FSA/AP produce the same strings but they assign different values to each string. Here it is shown that FSA/AP outperforms FSA alone by more accurately identifying a greater number of random strings and therefore missing fewer of them compared to the next level (CFG). Moreover, FSA only produces 12 different complexity values among all possible strings produced, while FSA/AP produces 34. The plot also shows that CFG is more powerful than LBA with halting runtime 27 and similar to LBA 54. The progression of LBA towards the full power of TM is also noticeable.}
\end{figure}

Running FSA is very fast; there is no halting problem and all stop very quickly. However, while FSA and CFG preserve some of the ranking of superior computational models and accelerate the appearance of `jumpers' (long strings with low complexity), these weak models of computation do not generate the strings with the highest algorithmic complexity that appear in the tail of the distributions of more powerful models, as shown in Fig.~\ref{missed}.

\begin{table}[!htb]
\centering
\begin{tabular}{rrrr}
  \hline
Size & Strings & Transducers \\ 
  \hline
8 & 256 &   1 \\ 
9 & 512 &   2 \\ 
10 & 1024 &   6 \\ 
11 & 2048 &  12 \\ 
12 & 4096 &  34 \\ 
13 & 8192 &  68 \\ 
14 & 16384 & 156 \\ 
15 & 32768 & 312 \\ 
16 & 65536 & 677 \\ 
17 & 131072 & 1354 \\ 
18 & 262144 & 2814 \\ 
19 & 524288 & 5628 \\ 
20 & 1048576 & 11474 \\ 
21 & 2097152 & 22948 \\ 
22 & 4194304 & 46332 \\
   \hline
\end{tabular}
\caption{Transducers per string length.} 
\label{table:validtrans} 
\end{table}

For example, of $2^8$ strings in $D_{S}(8)$, there is only one binary string of length 8 that encodes a transducer, which is the transducer with the transition function (\ref{transd:1}) (we refer to it as the smallest transducer).

In the case of $D_{S}(9)$, we found that out of $2^9$ strings, only two encode the smallest transducer with ``0'' and ``1'' as input. Again, the only string produced by this distribution is the empty string $\varepsilon$.

$D_{S}(16)$ is the first distribution in which one of the strings encodes a transducer with two states. The finite-state complexity $C_S(s)$ of the strings produced by $D_{S}(16)$ shows $8\leq C_S(s) \leq 12$ (see Table \ref{table:d16} and Table \ref{table2:d16} in the Supplementary Material).

$D_{S}(18)$ consists of 2814 transducers (see Table \ref{table:d18} and Table \ref{table2:d18} in the Supplementary Material). The longest string produced is of length 6. The rest of the tables are in the Supplementary Material.

\begin{table}[!ht]
\centering
\begin{tabular}{rr}
  \hline
String & Probability \\
  \hline
$\varepsilon$ & 0.82800 \\ 
  00 & 0.02701 \\ 
  11 & 0.02701 \\ 
  000 & 0.01990 \\ 
  111 & 0.01990 \\ 
  0 & 0.01848 \\ 
  1 & 0.01848 \\ 
  0000 & 0.01279 \\ 
  1111 & 0.01279 \\ 
  00000 & 0.00426 \\ 
  11111 & 0.00426 \\ 
  01 & 0.00213 \\ 
  10 & 0.00213 \\ 
  000000 & 0.00071 \\ 
  0101 & 0.00071 \\ 
  1010 & 0.00071 \\ 
  111111 & 0.00071 \\ 
   \hline
\end{tabular}
\caption{Example of a probability distribution for $D_{S}(18)$.} 
\label{table:d18}
\end{table}

\begin{table}[!ht]
\centering
\begin{tabular}{rr}
  \hline
Complexity & Frequency \\ 
  \hline
14 &   1024 \\ 
  12 &    640 \\ 
  13 &    512 \\ 
  11 &    320 \\ 
  10 &    212 \\ 
  9 &    106 \\ 
   \hline
\end{tabular}
\caption{Frequency of Finite-state complexity for strings produced by $D_{S}(18)$.} 
\label{table2:d18}
\end{table}

The rest of the tables are in the Supplementary Material.

\subsection{Computing finite-state complexity}

We performed another experiment in order to further analyze the characteristics of the finite-state complexity of all strings of length $n$. We summarize the results we obtained for computing finite-state complexity for each string $s$ of length $0 \leq n \leq 8$ in Table \ref{complexity:freq}, while Table \ref{complexity:trans} shows the strings $\sigma$ that encode transducers such that $T_{\sigma}^{S}(p)=s$. 

\begin{table}[ht]
\centering
\begin{tabular}{rr}
  \hline
 Complexity & Frequency \\ 
  \hline
4 &   1 \\ 
  7 &   2 \\ 
  8 &   2 \\ 
  9 &   4 \\ 
  10 &   4 \\ 
  11 &  10 \\ 
  12 &  22 \\ 
  13 &  32 \\ 
  14 &  56 \\ 
  15 & 126 \\ 
  16 & 252 \\ 
   \hline
\end{tabular}
\caption{Frequency of $C_{S}(s)$ for strings $s$ of length  $0 \leq n \leq 8$.} 
\label{complexity:freq}
\end{table}

\begin{table}[ht]
\centering
\begin{tabular}{rr}
  \hline
Transducer & Frequency \\ 
  \hline
0000 &   1 \\ 
  000100 &   8 \\ 
  0001010110 &   1 \\ 
  00010110 &   4 \\ 
  0001011100 &   1 \\ 
  0001011110 &   1 \\ 
  000110 &   8 \\ 
  00011100 &   4 \\ 
  0001110100 &   1 \\ 
  0001110110 &   1 \\ 
  0001111100 &   1 \\ 
  01000110 & 480 \\ 
   \hline
\end{tabular}
\caption{Frequency of the strings $\sigma$ that encode transducers such that $T_{\sigma}^{S}(p)=s$.} 
\label{complexity:trans}
\end{table}

\subsection{Context-free grammar distribution}

We created production rules for 298\,233 grammars with up to 26 non-terminal symbols and used the first 40\,000 of them to check how many of a set of 155 strings were produced by how many grammars for which we also had distribution numbers at all the other levelsa; FSA, LBA and TM. Table~\ref{cfgtable} shows the top 20 strings produced.

\begin{table}[!htb]
\centering
\begin{tabular}{rr}
  \hline
String & Frequency \\ 
  \hline
0 & 5131\\
00 & 5206\\
000 & 5536\\
0000 & 5480\\
00000 & 5508\\
000000 & 5508\\
00001 & 2818\\
0001 & 2810\\
00010 & 2754\\
00011 & 2764\\
001 & 2812\\
0010 & 2692\\
00100 & 2750\\
00101 & 2744\\
0011 & 2736\\
00110 & 2730\\
00111 & 2736\\
01 & 2688\\
010 & 2748\\
0100 & 2742\\
   \hline
\end{tabular}
\caption{Top 20 most frequent strings generated by context-free grammars (CFG). Assuming that the algorithmic Coding theorem held, we can then transform the frequency into a computable algorithmic complexity estimation by CFG for purposes of comparisoh.} 
\label{cfgtable} 
\end{table}

\subsection{Emergence of the Universal Distribution}

\begin{figure}[ht]
\includegraphics[width=12.3cm]{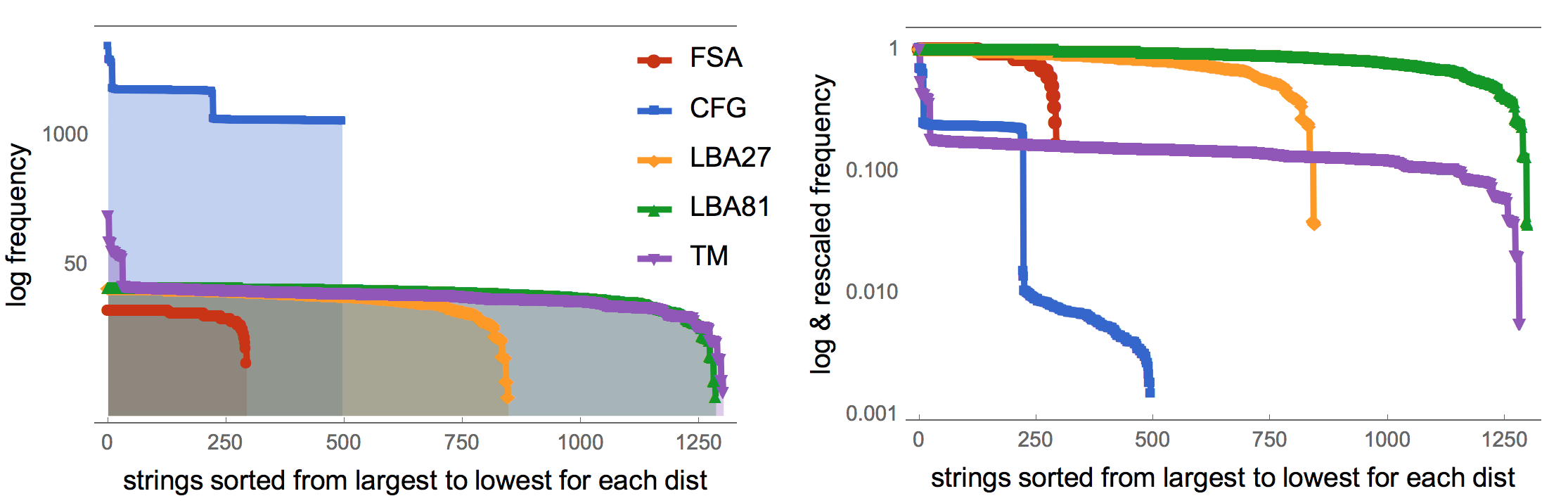} \caption{\label{comp}Comparison of empirical distributions of varying computational power, with TM in log plot (thus the only exponential) followed by LBA and then FSA and CFG when taking their fitting slope coefficients. Running CFG is very expensive because for every string we need to test whether it could be generated by all the grammars produced, which in our case means 1302 $\times$ 40\,000, and while the CYK algorithm runs in polynomial time, the combinatorial explosion renders the AP-based model by CFG intractable.}
\end{figure}

\subsubsection{Time-bounded Emergence}

\begin{figure}[ht]
\centering
\textbf{A}\hspace{6cm}\textbf{B}\\
\includegraphics[width=5.4cm]{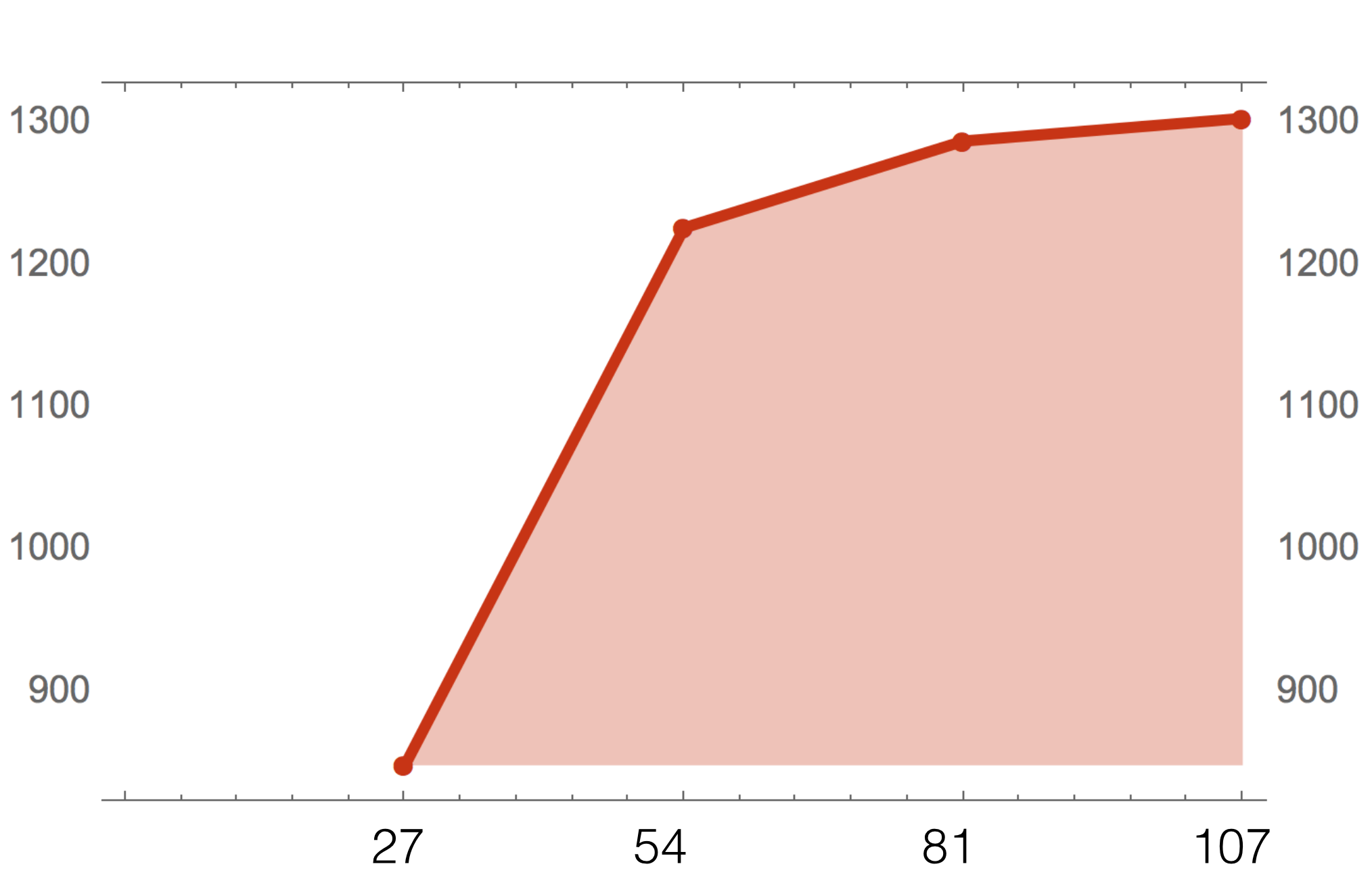}\hspace{.7cm}\includegraphics[width=5.4cm]{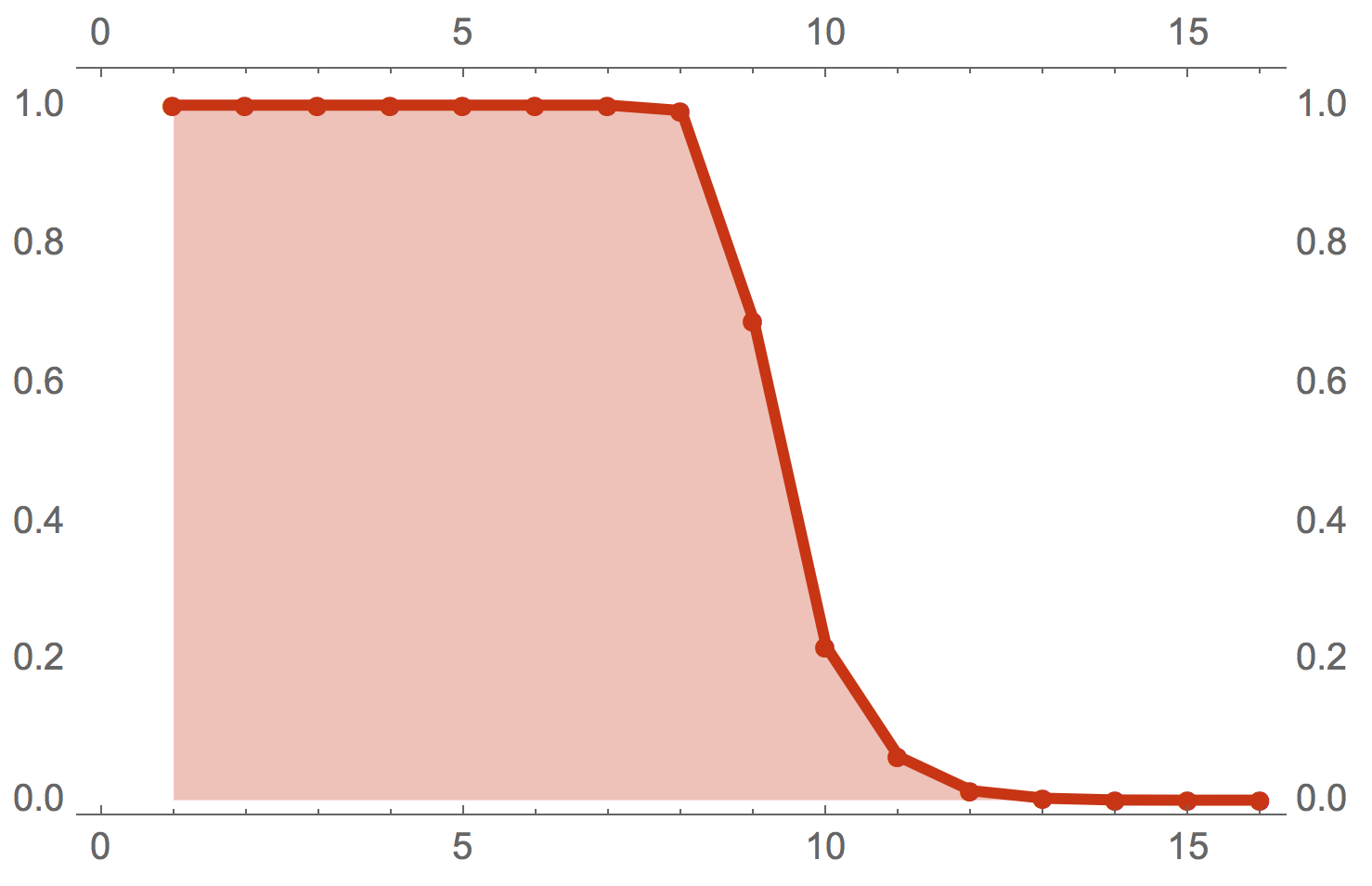}\\
\medskip

\textbf{C}\hspace{6cm}\textbf{D}\\
\includegraphics[width=5.5cm]{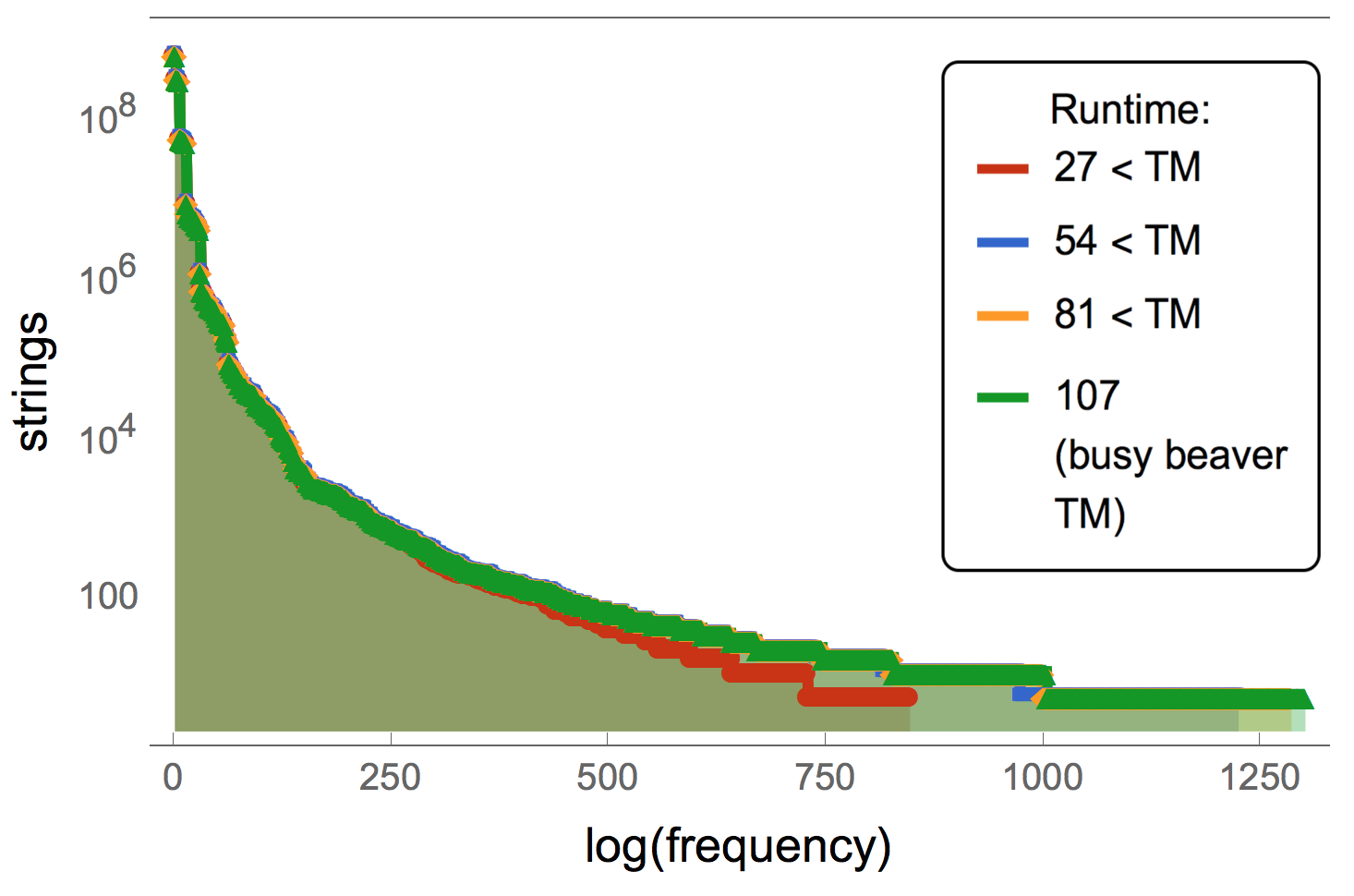}\hspace{.6cm}\includegraphics[width=5.5cm]{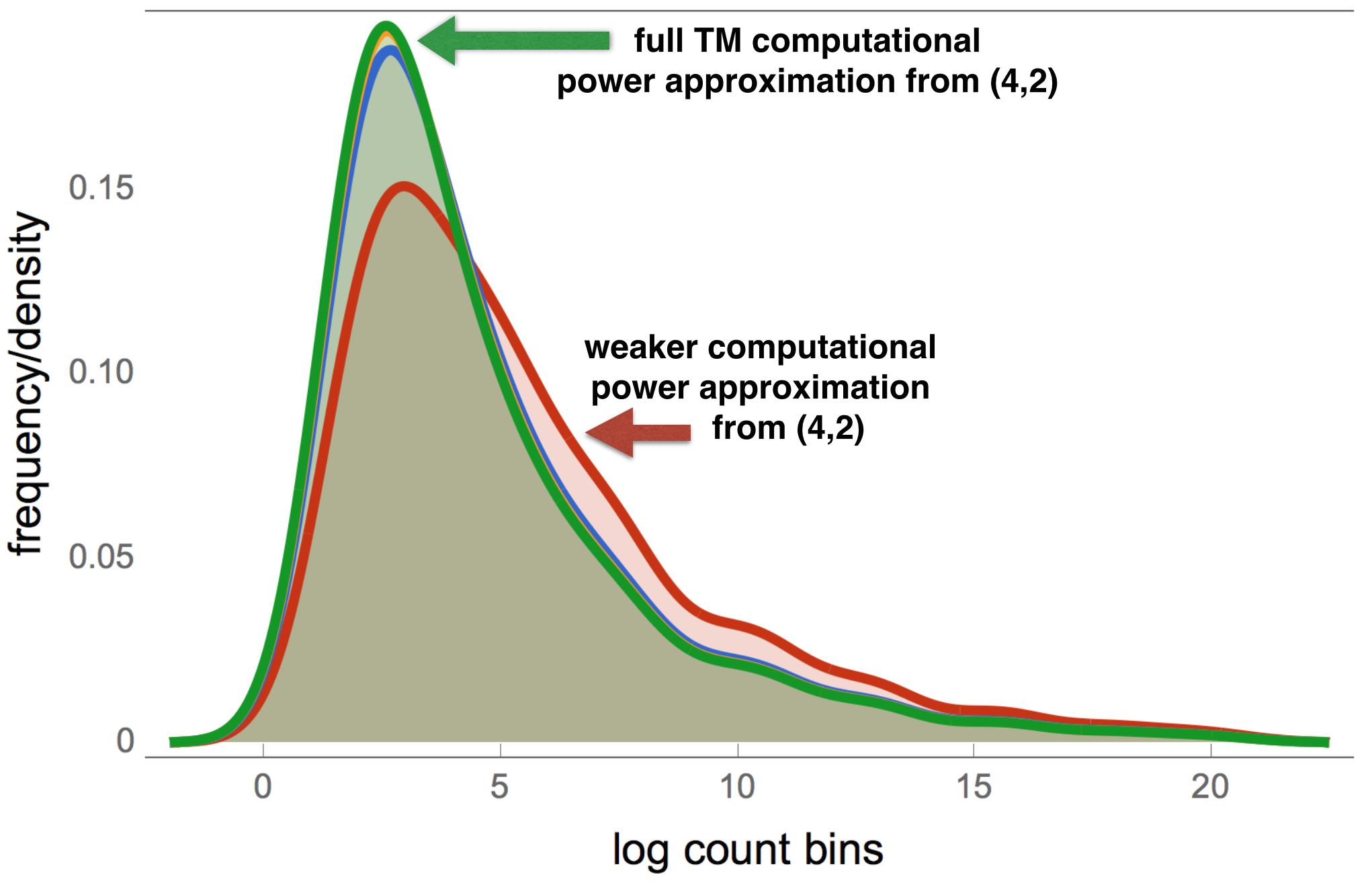}\\
 \caption{\label{timedist}Trading computational power for accuracy. (A) Number of halting machines per maximum imposed runtime (B) Number of strings produced among all possible, showing that all strings up to length 9 were generated in $(4,2)$ (C) Output frequency distribution per running time produced in $(4,2)$ from highest to lowest frequency (D) Smooth density plot maximizing the differences of the output distributions at each runtime approximating the full power of Turing machines in $(4,2)$ versus strictly lower computational power when bounding the runtime.}
\end{figure}

Fig.~\ref{timedist} shows how LBA asymptotically approximate the Universal Distribution.

As the results demonstrate (supported by Fig.~\ref{timedist}), by varying the allowed execution time for the space of Turing Machines we can approximate the CTM distribution corresponding to each level of the Chomsky hierarchy. For instance, regular languages (Type-3 grammars) can be decoded in \textit{linear time}, given that each transition and state in a finite automaton can be encoded by a corresponding state and transition in a Turing Machine. Context-free grammars (Type-2) can be decoded in \textit{polynomial time} with parsing algorithms such as CYK (Subsection~\ref{cyk}).

\subsubsection{Rate of convergence of the distributions}

\begin{figure}[ht!]
\centering
\includegraphics[width=10.5cm]{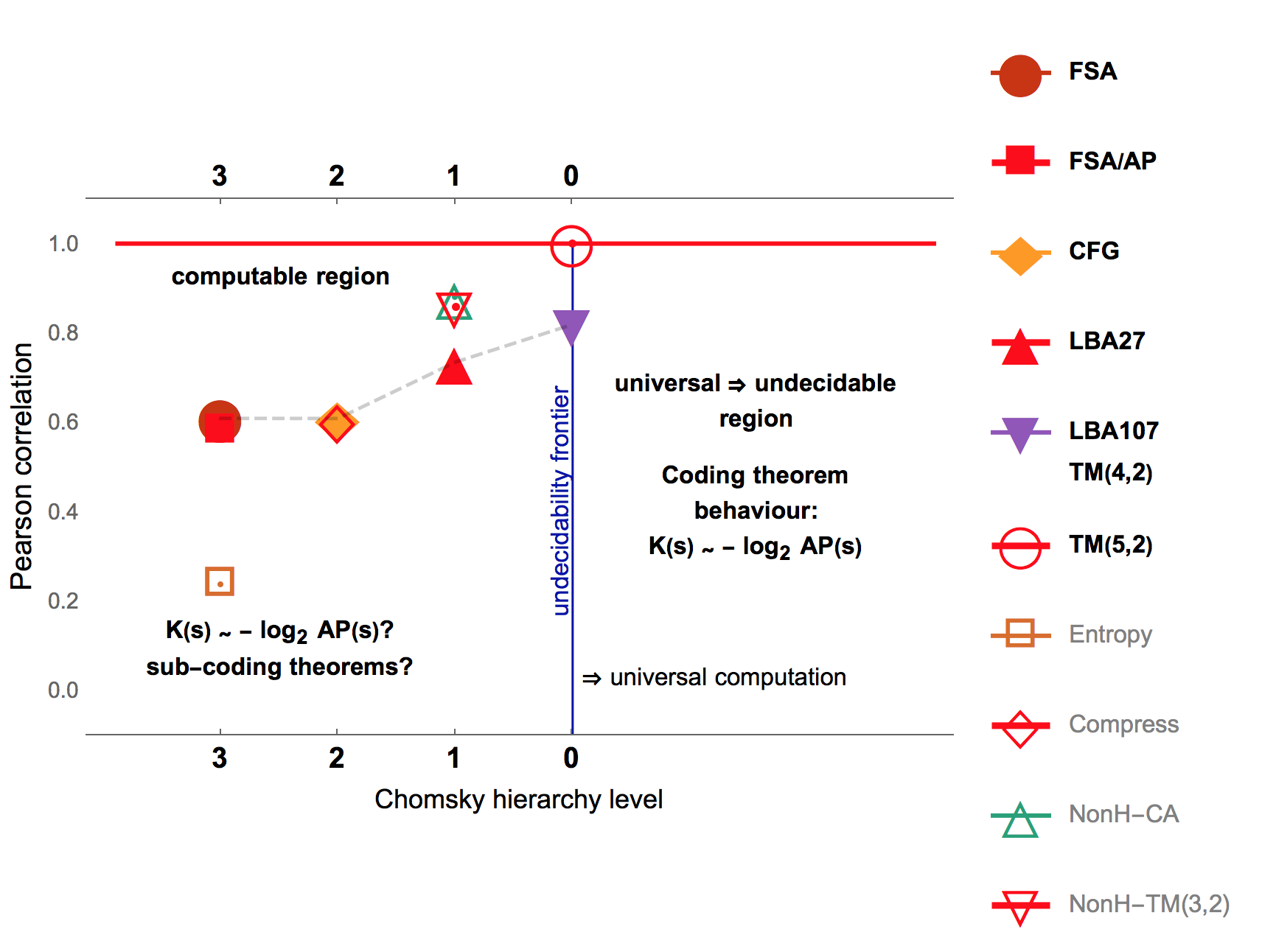}
 \caption{\label{mainfig}Coding theorem-like behaviour and emergence of the Universal Distribution. Correlation in rank (distributions were sorted in terms of each other) of empirical output distributions as compared to the output distribution of $TM(5,2)$ as introduced and thoroughly investigated in~\cite{solerzenil}. A progression towards greater correlation is noticed as a function of increasing computational power. Bold black labels are placed at their Chomsky level and gray labels are placed within the highest correlated level. Shannon entropy and lossless compression (Compress) distribute values below or at about the first 2 Chomsky types, as expected. It is not surprising to see the LBA with runtime 107 further deviate in ranking, because LBA after 27 steps produced the highest frequency strings, which are expected to converge faster. Eventually LBA 107 (which is none other than TM(4,2)) will converge to TM(5,2). An empirical bound of non-halting models seems to be low LBA even when increasing the number of states (or symbols for CA).}
\end{figure}

One opportunity afforded by this analysis is the assessment of the way in which other methods distribute strings by (statistical) randomness, such as Shannon entropy and other means of approximation of algorithmic complexity, such as lossless compression algorithms, in particular one of the most popular such methods, based on LZW (Compress). We can then compare these 2 methods in relation to estimations of a Universal Distribution produced by $TM(4,2)$. The results (see Fig.~\ref{mainfig}) of both entropy and Compression conform with the theoretical expectation. Entropy correlates best at the first level of the Chomsky hierarchy, that of FSA, stressing that the algorithmic discriminatory power of entropy to tell apart randomness from pseudo-randomness is limited to statistical regularities of the kind that regular languages would capture. Lossless compression, however, at least as assessed by one of the most popular methods underlying
other popular lossless compression formats, outperformed Shannon entropy, but not by much, and it was at best most closely correlated to the output distribution generated by CFG. This does not come as a surprise, given that popular implementations of lossless compression are a variation of Shannon entropy generalized to blocks (variable-width window) that capture repetitions, often followed by a remapping to use shorter codes for values with higher probabilities (dictionary encoding, Huffman coding)--hence effectively a basic grammar based on a simple rewriting rule system. We also found that, while non-halting models approximate the Universal Distribution, they start diverging from TM and remain correlated to LBA with lower runtimes, despite increasing the number of states. This may be expected from the over-representation of strings that non-halting machines would otherwise skip (defined as produced after halting for machines with halting configurations).

\section{Some open questions}

\subsection{Tighter bounds and suprema values}

We have provided upper and lower bounds for each model of computation, but current intervals seem to overlap for some measures of correlation. One question concerns the exact boundaries, especially how much closer to the Universal Distribution each supremum for each model of computation can take us, In other words, can they find the tighter bounds for the intervals at each level?

\subsection{Computable error corrections}

There are some very interesting open questions to explore in the future. For example, whether computable corrections can be made to subuniversal distributions, e.g. as calculated by context-free grammars, in order to correct for trivial (and also computable) biases such as string length. Indeed, while CFG produced an interesting distribution closer to that of LBA and better than FSA, there is, for example, an over-representation of trivial long strings which can easily be corrected. The suspicion is that while one can apply these corrections and increase the speed of convergence to TM, the interesting cases are non-computable. A theoretical framework and a numerical analysis would be interesting to develop. 

\subsection{Sensitivity to choice of enumeration}

A question equivalent to that of the choice of programming language of the reference universal Turing machine is the question of the stability of the chosen enumerations for the different models of computation, both at the same and at different computational power levels. In~\cite{calude} it is shown that FSA complexity is invariant to enumeration choice, and an open question is whether this happens at other levels of computational power between FSA and TM, e.g. CFG for some computational models of generative grammars (e.g. the one chosen in this paper). A striking outcome of the results reported here is that not only does the increase in computing power produce better approximations to the distribution produced by Turing machines, but that completely different models of computation---not only in terms of language description but also in terms of computational power--- produce similar distributions, thus suggesting a greater independence of the choice of arbitrary parameters, such as enumeration or model. In~\cite{algonature} we found such convergence in values and ranking in experiments comparing Turing machines, cellular automata and Post tag systems, and moreover these results suggested some sort of `natural behaviour' defined as behaviour that is not artificially introduced with the purpose of producing a different-looking initial distribution (before converging per the invariance theorem for systems with such a property). 

Thus in~\cite{fundamentainformaticapaper} we proposed a measure of `algorithmicity' in the sense of the Universal Distribution, quantifying how close or removed a method producing a distribution is from other approximations, in particular that of one of the most standard Turing machine models, the one used for the Busy Beaver~\cite{rado}, that we have shown is not a special case-- several other variations of this model and completely different models of computation produce similar output distributions~\cite{kolmo2d,computabilitypaper,fundamentainformaticapaper}, including the results reported in this paper. However, one other open question would entail enumerating systems in different ways and numerically quantifying how many of them are convergent or divergent, to what degree the divergent diverge and for how long and under what conditions, and whether the convergent dominate.

\subsection{Complexity of missing strings from non-halting models}

We have shown that for halting models of computation, decreasing the power of the computational model has the effect of missing
some of the most algorithmically random strings that are produced in the model with the next highest computational power. As we have also seen, non-halting models seem to converge to lower runtime distributions (LBA, which even when they are highly correlated to TMs, do not appear to asymptotically approach TMs but to remain closer to LBA. An interesting question to explore concerns the complexity of the strings missed by non-halting machines or which strings are under or over-represented. As opposed to the strings missed by halting machines below Turing universality, models of computation without halting configurations but still capable of Turing universality (defined on non-halting models) may miss different kinds of strings than those models with halting configurations but strictly weaker computational power (e.g. LBA). Do they miss more or fewer random strings as compared to halting models?

\section{Conclusions}

It is interesting to explore and seek \textit{algorithmic Coding-theorem} like behaviour in subuniversal systems to better understand the emergence, and rate of emergence, of properties related to algorithmic probability, in particular those established by the \textit{algorithmic Coding-theorem} from the Universal Distribution--and thus of algorithmic complexity--for types of computational systems of different computational power, in particular below Turing universality.

The results reported show that the closer a system is to Turing-completeness, the closer its output distribution is to the empirical distribution produced by a complete set of small Turing machines, and that finite approximations of algorithmic complexity from subuniversal systems~\cite{solerzenilcomplexity} produce a gradual emergence of properties related to the Universal Distribution.

The results also show improvements over current major tools for approximating algorithmic complexity, such as lossless compression algorithms. To our knowledge, it has not been possible to date to quantify or compare the performance of lossless compression algorithms as estimations of algorithmic complexity, as there was no other standard or alternative. The construction of empirical distributions does provide the means and constitutes an approach to evaluate performance in what we have named an \textit{algorithmicity test}~\cite{algonature}.

Compared to entropic and the most popular lossless compression algorithms, approximations to algorithmic complexity from finite approximations of Algorithmic Probability--even over weak models of computation--constitutes a major improvement over strings that are otherwise assigned a greater randomness content by traditional methods such as common lossless compression, Shannon entropy and equivalent statistical formulations.

This algorithmic behavioural exploration is particularly relevant because processes in nature may be weaker than the full power of Turing machines, due to access to finite resources and because of physical degradation. For example, some biological and genetic processes such as DNA translation and RNA transcription can only decode and produce messages and sequences by reading in a single-strand direction and thus may be modelled by Context-free grammars (CFG). The results here suggest that, for example, RNA transcription would not sample uniformly the whole aminoacid space, but would favour the production of simple protein primary structures. In this way the approximation to the Universal Distribution is not only descriptive but does play an active role in active (e.g. biological) processes. Given these types of questions, we also explored non-halting models, as natural processes may not have definite and distinguishable halting states, unlike, e.g., RNA transcription that, exceptionally, has a clear halting condition represented by the RNA polymerase enzyme reading the stop codons.

We have shown that up to 60\% of the simplicity vs complexity bias produced even by the weakest of the computational models (FST/FSA) in the Chomsky hierarchy (Type-3)--disregarding halting configuration--can be explained by or accounted for by Algorithmic Probability in the model output approximation to the Universal Distribution as tested by empirical distributions compared to the empirical distributions produced by large sets of small Turing machines. On the one hand, while producing empirical distributions from FSA is computationally cheap and produces a large set of strings mostly distributed in the shape of the Universal Distribution, it produces very few strings with different probability/complexity values and thus provides very little discriminatory power. On the other hand, producing strings at the level of CFG (Type-2) with grammars turned out to be computationally very expensive, despite the existence of a polynomial algorithm that can verify string production. However, when a set of strings needs to be checked against a set of grammars, the computational task becomes exponential by a combinatorial argument. Type-2, however, can also be simulated by not only limiting the tape length in LBAs but also forcing them to operate in only one tape direction, which would lead to a simulation similar to that of LBAs. Finally, we have demonstrated that LBAs (Type-1) are both the most direct and the fastest computational model approximating the empirical distributions produced by Turing machines (Type-0) that we generated.

\section*{Acknowledgements}

L.B and F.H.-Q. were supported by the grant No. 221341 from SEP-Conacyt
CB-2013-01, and H.Z. by grant No. 2015-05299 from the Swedish Research Council (Vetenskapsr\r{a}det).

\newpage

\section*{Supplementary Material}

\subsubsection{Empirical distributions $<D_{S}(19)$ for FSA}

For $D_{S}(10)$ we have it that there are six strings that encode a transducer. In fact, two of them are different from the smallest transducer. However, as in the previous cases, the only string produced is the empty string $\varepsilon$.

$D_{S}(11)$ consists of 12 strings with inputs of length one and three but the output of these transducers is $\varepsilon$.

$D_{S}(12)$ contains 34 transducers with inputs whose lengths range from two to four. The output distribution still contains only the string $\varepsilon$.

$D_{S}(13)$ is a more interesting distribution. It consists of 68 strings whose input has lengths 1, 3 and 5. Table \ref{table:d13} shows the probability distribution of the strings produced by this distribution. The finite-state complexity of the strings that comprise $D_{S}(13)$ is summarized in Table \ref{table2:d13}.

\begin{table}[!ht]
\centering
\begin{tabular}{rr}
  \hline
 String & Probability \\ 
  \hline
$\varepsilon$ & 0.94118 \\ 
  0 & 0.02941 \\ 
  1 & 0.02941 \\ 
   \hline
\end{tabular}
\caption{Probability distribution of $D_{S}(13)$.} 
\label{table:d13}
\end{table}

\begin{table}[!ht]
\centering
\begin{tabular}{rr}
  \hline
Complexity & Frequency \\ 
  \hline
9 &     32 \\ 
  7 &     20 \\ 
  8 &     16 \\ 
   \hline
\end{tabular}
\caption{Frequency of finite-state complexity for strings produced by $D_{S}(13)$.} 
\label{table2:d13}
\end{table}

$D_{S}(14)$ is a richer distribution than the previous one since it contains 156 strings that encode different transducers. Table \ref{table:d14} shows the different strings produced by this distribution.

\begin{table}[!ht]
\centering
\begin{tabular}{rr}
  \hline
 String & Probability \\
  \hline
$\varepsilon$ & 0.92308 \\ 
  0 & 0.02564 \\ 
  1 & 0.02564 \\ 
  00 & 0.01282 \\ 
  11 & 0.01282 \\ 
   \hline
\end{tabular}
\caption{Probability distribution of $D_{S}(14)$.} 
\label{table:d14}
\end{table}

We note the following facts:

\begin{myitemize}
\item The length of the longest string produced is two.
\item The string $\varepsilon$ remains the one with the highest probability.
\item The finite-state complexity of the strings produced ranges from 7 to 10 (see Table \ref{table2:d14}).
\item $D_{S}(14)$ produces two strings of length 2 out of $2^2$, that is, ``00'' and ``11''. 
\end{myitemize}

\begin{table}[!ht]
\centering
\begin{tabular}{rr}
  \hline
Complexity & Frequency \\ 
  \hline
10 &     64 \\ 
  8 &     40 \\ 
  9 &     32 \\ 
  7 &     20 \\ 
   \hline
\end{tabular}
\caption{Frequency of Finite-state complexity for strings produced by $D_{S}(14)$.} 
\label{table2:d14}
\end{table}

$D_{S}(15)$ is quite similar to $D_{S}(14)$ in terms of the strings it produces (see Table \ref{table:d14} and Table \ref{table2:d15}). 

\begin{table}[!ht]
\centering
\begin{tabular}{rr}
  \hline
 String & Probability \\
  \hline
$\varepsilon$ & 0.88462 \\ 
  0 & 0.03205 \\ 
  1 & 0.03205 \\ 
  00 & 0.01923 \\ 
  11 & 0.01923 \\ 
  000 & 0.00641 \\ 
  111 & 0.00641 \\ 
   \hline
\end{tabular}
\caption{Probability distribution of $D_{S}(15)$.} 
\label{table:d15}
\end{table}

\begin{table}[!ht]
\centering
\begin{tabular}{rr}
  \hline
Complexity & Frequency \\ 
  \hline
11 &    128 \\ 
  9 &     80 \\ 
  10 &     64 \\ 
  8 &     40 \\ 
   \hline
\end{tabular}
\caption{Frequency of Finite-state complexity for strings produced by $D_{S}(15)$.} 
\label{table2:d15}
\end{table}

\begin{table}[!ht]
\centering
\begin{tabular}{rr}
  \hline
 String & Probability \\
  \hline
$\varepsilon$ & 0.87592 \\ 
  0 & 0.02363 \\ 
  00 & 0.02363 \\ 
  1 & 0.02363 \\ 
  11 & 0.02363 \\ 
  000 & 0.01182 \\ 
  111 & 0.01182 \\ 
  0000 & 0.00295 \\ 
  1111 & 0.00295 \\ 
   \hline
\end{tabular}
\caption{Probability distribution of $D_{S}(16)$.} 
\label{table:d16}
\end{table}

\begin{table}[!ht]
\centering
\begin{tabular}{rr}
  \hline
Complexity & Frequency \\ 
  \hline
12 &    256 \\ 
  10 &    160 \\ 
  11 &    128 \\ 
  9 &     80 \\ 
  8 &     53 \\ 
   \hline
\end{tabular}
\caption{Frequency of Finite-state complexity for strings produced by $D_{S}(16)$.} 
\label{table2:d16}
\end{table}

$D_{S}(17)$ shows an even more diverse set of strings produced (see Table \ref{table:d17}). We have the following interesting facts,

\begin{myitemize}
\item The longest string produced is of length 5.
\item For the first time, a distribution produces all strings of length 2.
\end{myitemize}

\begin{table}[!ht]
\centering
\begin{tabular}{rr}
  \hline
 String & Probability \\
  \hline
$\varepsilon$ & 0.83752 \\ 
  0 & 0.02806 \\ 
  1 & 0.02806 \\ 
  00 & 0.02511 \\ 
  11 & 0.02511 \\ 
  000 & 0.01773 \\ 
  111 & 0.01773 \\ 
  0000 & 0.00739 \\ 
  1111 & 0.00739 \\ 
  00000 & 0.00148 \\ 
  01 & 0.00148 \\ 
  10 & 0.00148 \\ 
  11111 & 0.00148 \\ 
   \hline
\end{tabular}
\caption{Probability distribution of $D_{S}(17)$.} 
\label{table:d17}
\end{table}

\begin{table}[!ht]
\centering
\begin{tabular}{rr}
  \hline
Complexity & Frequency \\ 
  \hline
13 &    512 \\ 
  11 &    320 \\ 
  12 &    256 \\ 
  10 &    160 \\ 
  9 &    106 \\ 
   \hline
\end{tabular}
\caption{Frequency of finite-state complexity for strings produced by $D_{S}(17)$.} 
\label{table2:d17}
\end{table}

\subsubsection{$D_{S}(19)$,$D_{S}(20)$,$D_{S}(21)$ and $D_{S}(22)$}

Here are the strings that comprise each one of these distributions.   

\begin{table}[!ht]
\centering
\begin{tabular}{rr}
  \hline
 String & Probability \\
  \hline
$\varepsilon$ & 0.80597 \\ 
  000 & 0.02345 \\ 
  111 & 0.02345 \\ 
  00 & 0.02274 \\ 
  11 & 0.02274 \\ 
  0 & 0.01812 \\ 
  1 & 0.01812 \\ 
  0000 & 0.01706 \\ 
  1111 & 0.01706 \\ 
  00000 & 0.00817 \\ 
  11111 & 0.00817 \\ 
  000000 & 0.00284 \\ 
  111111 & 0.00284 \\ 
  01 & 0.00178 \\ 
  10 & 0.00178 \\ 
  0101 & 0.00107 \\ 
  1010 & 0.00107 \\ 
  0000000 & 0.00036 \\ 
  001 & 0.00036 \\ 
  010 & 0.00036 \\ 
  010101 & 0.00036 \\ 
  011 & 0.00036 \\ 
  100 & 0.00036 \\ 
  101 & 0.00036 \\ 
  101010 & 0.00036 \\ 
  110 & 0.00036 \\ 
  1111111 & 0.00036 \\ 
   \hline
\end{tabular}
\caption{Probability distribution of $D_{S}(19)$. `Jumpers' as defined in~\cite{zenild4} and~\cite{solerzenil} are apparent, those simple strings of relatively much greater length that climb the complexity ladder.} 
\label{table:d19}
\end{table}

\begin{table}[!ht]
\centering
\begin{tabular}{rr}
  \hline
Complexity & Frequency \\ 
  \hline
15 &   2048 \\ 
  13 &   1280 \\ 
  14 &   1024 \\ 
  12 &    640 \\ 
  11 &    424 \\ 
  10 &    212 \\ 
   \hline
\end{tabular}
\caption{Frequency of Finite-state complexity for strings produced by $D_{S}(19)$.} 
\label{table2:d19}
\end{table}
 
\begin{table}[!ht]
\centering
\begin{tabular}{rr}
  \hline
String & Probability \\ 
  \hline
$\varepsilon$ & 0.80024 \\ 
  0000 & 0.02144 \\ 
  1111 & 0.02144 \\ 
  00 & 0.02092 \\ 
  000 & 0.02092 \\ 
  11 & 0.02092 \\ 
  111 & 0.02092 \\ 
  0 & 0.01185 \\ 
  00000 & 0.01185 \\ 
  1 & 0.01185 \\ 
  11111 & 0.01185 \\ 
  000000 & 0.00593 \\ 
  111111 & 0.00593 \\ 
  01 & 0.00174 \\ 
  10 & 0.00174 \\ 
  0101 & 0.00157 \\ 
  1010 & 0.00157 \\ 
  0000000 & 0.00139 \\ 
  1111111 & 0.00139 \\ 
  010101 & 0.00070 \\ 
  101010 & 0.00070 \\ 
  00000000 & 0.00035 \\ 
  11111111 & 0.00035 \\ 
  0001 & 0.00017 \\ 
  0010 & 0.00017 \\ 
  0011 & 0.00017 \\ 
  0100 & 0.00017 \\ 
  01010101 & 0.00017 \\ 
  0110 & 0.00017 \\ 
  0111 & 0.00017 \\ 
  1000 & 0.00017 \\ 
  1001 & 0.00017 \\ 
  10101010 & 0.00017 \\ 
  1011 & 0.00017 \\ 
  1100 & 0.00017 \\ 
  1101 & 0.00017 \\ 
  1110 & 0.00017 \\ 
   \hline
\end{tabular}
\caption{Probability distribution of $D_{S}(20)$.} 
\label{table:d20d}
\end{table}

\begin{table}[!ht]
\centering
\begin{tabular}{rr}
  \hline
Complexity & Frequency \\ 
  \hline
16 &   4096 \\ 
  14 &   2560 \\ 
  15 &   2048 \\ 
  13 &   1280 \\ 
  12 &    848 \\ 
  11 &    424 \\ 
  10 &    218 \\ 
   \hline
\end{tabular}
\caption{Frequency of finite-state complexity for strings produced by $D_{S}(20)$.} 
\label{table2:d20f}
\end{table}

\begin{longtable}{rr}
  \hline
String & Probability \\ 
  \hline
$\varepsilon$ & 0.78630 \\ 
  0000 & 0.02109 \\ 
  1111 & 0.02109 \\ 
  000 & 0.02066 \\ 
  111 & 0.02066 \\ 
  00 & 0.01682 \\ 
  11 & 0.01682 \\ 
  00000 & 0.01665 \\ 
  11111 & 0.01665 \\ 
  0 & 0.01063 \\ 
  1 & 0.01063 \\ 
  000000 & 0.00959 \\ 
  111111 & 0.00959 \\ 
  0000000 & 0.00331 \\ 
  1111111 & 0.00331 \\ 
  01 & 0.00166 \\ 
  10 & 0.00166 \\ 
  0101 & 0.00139 \\ 
  1010 & 0.00139 \\ 
  00000000 & 0.00122 \\ 
  11111111 & 0.00122 \\ 
  010101 & 0.00105 \\ 
  101010 & 0.00105 \\ 
  01010101 & 0.00044 \\ 
  10101010 & 0.00044 \\ 
  001 & 0.00026 \\ 
  010 & 0.00026 \\ 
  011 & 0.00026 \\ 
  100 & 0.00026 \\ 
  101 & 0.00026 \\ 
  110 & 0.00026 \\ 
  000000000 & 0.00009 \\ 
  0000000000 & 0.00009 \\ 
  00001 & 0.00009 \\ 
  00010 & 0.00009 \\ 
  00011 & 0.00009 \\ 
  00100 & 0.00009 \\ 
  00101 & 0.00009 \\ 
  00110 & 0.00009 \\ 
  00111 & 0.00009 \\ 
  01000 & 0.00009 \\ 
  01001 & 0.00009 \\ 
  01010 & 0.00009 \\ 
  0101010101 & 0.00009 \\ 
  01011 & 0.00009 \\ 
  01100 & 0.00009 \\ 
  01101 & 0.00009 \\ 
  01110 & 0.00009 \\ 
  01111 & 0.00009 \\ 
  10000 & 0.00009 \\ 
  10001 & 0.00009 \\ 
  10010 & 0.00009 \\ 
  10011 & 0.00009 \\ 
  10100 & 0.00009 \\ 
  10101 & 0.00009 \\ 
  1010101010 & 0.00009 \\ 
  10110 & 0.00009 \\ 
  10111 & 0.00009 \\ 
  11000 & 0.00009 \\ 
  11001 & 0.00009 \\ 
  11010 & 0.00009 \\ 
  11011 & 0.00009 \\ 
  11100 & 0.00009 \\ 
  11101 & 0.00009 \\ 
  11110 & 0.00009 \\ 
  111111111 & 0.00009 \\ 
  1111111111 & 0.00009 \\ 
   \hline
\hline
\caption{Probability distribution of $D_{S}(21)$. } 
\label{table:d21}
\end{longtable}

\begin{table}[!ht]
\centering
\begin{tabular}{rr}
  \hline
Complexity & Frequency \\ 
  \hline
17 &   8192 \\ 
  15 &   5120 \\ 
  16 &   4096 \\ 
  14 &   2560 \\ 
  13 &   1696 \\ 
  12 &    848 \\ 
  11 &    436 \\ 
   \hline
\end{tabular}
\caption{Frequency of Finite-state complexity for strings produced by $D_{S}(21)$.} 
\label{table2:d20}
\end{table}

\paragraph{}
\mbox{}
\begin{longtable}{rr}
  \hline
String & Probability \\ 
  \hline
$\varepsilon$ & 0.78313 \\ 
  0000 & 0.02180 \\ 
  1111 & 0.02180 \\ 
  000 & 0.01744 \\ 
  111 & 0.01744 \\ 
  00000 & 0.01727 \\ 
  11111 & 0.01727 \\ 
  000000 & 0.01442 \\ 
  111111 & 0.01442 \\ 
  00 & 0.01416 \\ 
  11 & 0.01416 \\ 
  0 & 0.00622 \\ 
  1 & 0.00622 \\ 
  0000000 & 0.00587 \\ 
  1111111 & 0.00587 \\ 
  00000000 & 0.00276 \\ 
  11111111 & 0.00276 \\ 
  0101 & 0.00160 \\ 
  1010 & 0.00160 \\ 
  01 & 0.00138 \\ 
  10 & 0.00138 \\ 
  010101 & 0.00125 \\ 
  101010 & 0.00125 \\ 
  01010101 & 0.00073 \\ 
  10101010 & 0.00073 \\ 
  000000000 & 0.00043 \\ 
  111111111 & 0.00043 \\ 
  0000000000 & 0.00030 \\ 
  1111111111 & 0.00030 \\ 
  0101010101 & 0.00026 \\ 
  1010101010 & 0.00026 \\ 
  001 & 0.00017 \\ 
  010 & 0.00017 \\ 
  011 & 0.00017 \\ 
  100 & 0.00017 \\ 
  101 & 0.00017 \\ 
  110 & 0.00017 \\ 
  0001 & 0.00009 \\ 
  0010 & 0.00009 \\ 
  001001 & 0.00009 \\ 
  0011 & 0.00009 \\ 
  0100 & 0.00009 \\ 
  010010 & 0.00009 \\ 
  0110 & 0.00009 \\ 
  011011 & 0.00009 \\ 
  0111 & 0.00009 \\ 
  1000 & 0.00009 \\ 
  1001 & 0.00009 \\ 
  100100 & 0.00009 \\ 
  1011 & 0.00009 \\ 
  101101 & 0.00009 \\ 
  1100 & 0.00009 \\ 
  1101 & 0.00009 \\ 
  110110 & 0.00009 \\ 
  1110 & 0.00009 \\ 
  000000000000 & 0.00004 \\ 
  000001 & 0.00004 \\ 
  000010 & 0.00004 \\ 
  000011 & 0.00004 \\ 
  000100 & 0.00004 \\ 
  000101 & 0.00004 \\ 
  000110 & 0.00004 \\ 
  000111 & 0.00004 \\ 
  001000 & 0.00004 \\ 
  001010 & 0.00004 \\ 
  001011 & 0.00004 \\ 
  001100 & 0.00004 \\ 
  001101 & 0.00004 \\ 
  001110 & 0.00004 \\ 
  001111 & 0.00004 \\ 
  010000 & 0.00004 \\ 
  010001 & 0.00004 \\ 
  010011 & 0.00004 \\ 
  010100 & 0.00004 \\ 
  010101010101 & 0.00004 \\ 
  010110 & 0.00004 \\ 
  010111 & 0.00004 \\ 
  011000 & 0.00004 \\ 
  011001 & 0.00004 \\ 
  011010 & 0.00004 \\ 
  011100 & 0.00004 \\ 
  011101 & 0.00004 \\ 
  011110 & 0.00004 \\ 
  011111 & 0.00004 \\ 
  100000 & 0.00004 \\ 
  100001 & 0.00004 \\ 
  100010 & 0.00004 \\ 
  100011 & 0.00004 \\ 
  100101 & 0.00004 \\ 
  100110 & 0.00004 \\ 
  100111 & 0.00004 \\ 
  101000 & 0.00004 \\ 
  101001 & 0.00004 \\ 
  101010101010 & 0.00004 \\ 
  101011 & 0.00004 \\ 
  101100 & 0.00004 \\ 
  101110 & 0.00004 \\ 
  101111 & 0.00004 \\ 
  110000 & 0.00004 \\ 
  110001 & 0.00004 \\ 
  110010 & 0.00004 \\ 
  110011 & 0.00004 \\ 
  110100 & 0.00004 \\ 
  110101 & 0.00004 \\ 
  110111 & 0.00004 \\ 
  111000 & 0.00004 \\ 
  111001 & 0.00004 \\ 
  111010 & 0.00004 \\ 
  111011 & 0.00004 \\ 
  111100 & 0.00004 \\ 
  111101 & 0.00004 \\ 
  111110 & 0.00004 \\ 
  111111111111 & 0.00004 \\ 
   \hline
\hline
\caption{Probability distribution of $D_{S}(22)$.} 
\label{table:d22}
\end{longtable}
\paragraph{}
\mbox{}
\begin{table}[!ht]
\centering
\begin{tabular}{rr}
  \hline
Complexity & Frequency \\ 
  \hline
18 &  16384 \\ 
  16 &  10240 \\ 
  17 &   8192 \\ 
  15 &   5120 \\ 
  14 &   3392 \\ 
  13 &   1696 \\ 
  12 &    872 \\ 
  11 &    436 \\ 
   \hline
\end{tabular}
\caption{Frequency of Finite-state complexity for strings produced by $D_{S}(22)$.} 
\label{table2:d20f2}
\end{table}
\paragraph{}
\mbox{}

\subsubsection{Code in Python for finite-state complexity}

The program distributionTransducers.py is used to analyze all strings $\tau$ of  length $|\tau|=n$ to determine whether $\tau$ satisfies  $\tau = \sigma^\frown p$ (for $\sigma \in S$), and if so then the program computes $T_{\sigma}^{S}(p)$. This program generates a set of output strings (result-experiment-distribution.csv) from which we can construct an output frequency distribution.

Example of execution:

\begin{myitemize}
\item python distributionTransducers.py 8 10 analyzes all strings of length 8 up to length 10.
\item python distributionTransducers.py 8 8 analyzes all strings of length 8.
\end{myitemize}  

The file result-experiment-distribution.csv contains the following columns:

\begin{myitemize}
\item string: this corresponds to the strings $\tau$ discussed above.
\item valid-encoding: takes value 1 in case $\tau = \sigma^\frown p$ and 0 otherwise.
\item sigma: corresponds to string $\sigma$ such that $\tau = \sigma^\frown p$.
\item string-p: corresponds to string $p$ such that $T_{\sigma}^{S}(p)=s$.
\item num-states: number of states of transducer $T_{\sigma}^{S}(p)$.
\item output: corresponds to string $s$ such that $T_{\sigma}^{S}(p)=s$.
\item output-complexity: finite-state complexity of output string $x$.
\end{myitemize}

The program computeComplexityStrings.py computes the finite-state complexity for all strings of length $n$ up to length $m$ (this is the implementation of the algorithm described in~\cite{calude}). This program generates the file result-complexity.csv which contains the following columns:

\begin{myitemize}
\item $s$: the string that the program is calculating the finite-state complexity for.
\item complexity: finite-state complexity of string $s$.
\item sigma: string $\sigma$ such that $T_{\sigma}^{S}(p)=s$.
\item string-p, string $p$ such that $T_{\sigma}^{S}(p)=s$.
\end{myitemize}


\begin{thebibliography}{99}

\bibitem{antunes} L. Antunes y L. Fortnow, Time-Bounded Universal Distributions, Electronic Colloquium on Computational Complexity, Report No. 144, 2005.
\bibitem{bienvenu} L. Bienvenu and R. Downey. \textit{Kolmogorov complexity and Solovay
functions}. In STACS, volume 3 of LIPIcs, pages 147--158. Schloss Dagstuhl-
Leibniz-Zentrum fuer Informatik, 2009.
\bibitem{campeanu} C. Campeanu, K. Culik II, K. Salomaa, S. Yu, State complexity of basic operations on finite languages. In: O. Boldt, H. J\"urgensen, (eds.) \textit{WIA 1999}, LNCS, vol. 2214, pp. 60--70. Springer, Heidelberg, 2001.
\bibitem{fundamentainformaticapaper} J.-P. Delahaye, H. Zenil, Towards a stable definition of Kolmogorov-Chaitin complexity, arXiv:0804.3459 [cs.IT], 2008.
\bibitem{zenild4} J.-P. Delahaye and H. Zenil, Numerical Evaluation of the Complexity of Short Strings: A Glance Into the Innermost Structure of Algorithmic Randomness, \textit{Applied Mathematics and Computation} 219, pp. 63--77, 2012.
\bibitem{kamal1}K. Dingle, S. Schaper, and A.A. Louis, The structure of the genotype-phenotype map strongly constrains the evolution of non-coding RNA, \textit{Interface Focus} 5: 20150053, 2015.
\bibitem{gauvrit2} N. Gauvrit, F. Soler-Toscano, H. Zenil, Natural Scene Statistics Mediate the Perception of Image Complexity, \textit{Visual Cognition,} vol. 22:8, pp. 1084--1091, 2014.
\bibitem{rado} T. Rado, ``On non-computable functions'' \textit{Bell System Technical Journal} 41:3, 877--884, 1962.
\bibitem{ploscompbio} N. Gauvrit, H. Zenil, F. Soler-Toscano, J.-P. Delahaye, P. Brugger, Human Behavioral Complexity Peaks at Age 25, \textit{PLoS Comput Biol} 13(4): e1005408, 2017.
\bibitem{kamal2} S.F. Greenbury, I.G. Johnston, A.A. Louis, S.E. Ahnert, J. R., A tractable genotype-phenotype map for the self-assembly of protein quaternary structure, \textit{Soc. Interface} 11, 20140249, 2014.
\bibitem{evocomplex} S. Hern\'andez-Orozco, H. Zenil, N.A. Kiani, Algorithmically probable mutations reproduce aspects of evolution such as convergence rate, genetic memory, modularity, diversity explosions, and mass extinction, arXiv:1709.00268 [cs.NE]
\bibitem{kempe} V. Kempe, N. Gauvrit, D. Forsyth, Structure emerges faster during cultural transmission in children than in adults, \textit{Cognition}, 136, 247--254, 2014.
\bibitem{holzl} R. H\"olzl, T. Kr\"aling, W. Merkle, Time-bounded Kolmogorov complexity and Solovay functions, \textit{Theory Comput. Syst.} 52:1, 80--94, 2013.
\bibitem{AIXI} M. Hutter, \textit{A Theory of Universal Artificial Intelligence based on Algorithmic Complexity}, Springer, 2000.
\bibitem{levin} L.A Levin, Universal sequential search problems. \textit{Problems of Information Transmission}, 9:265--266, 1973.
\bibitem{levin1974} L.A. Levin. Laws of information conservation (non-growth) and aspects of the foundation of probability theory, \textit{Problems Information Transmission,} 10(3):206--210, 1974.
\bibitem{kraft} L.G. Kraft, A device for quantizing, grouping, and coding amplitude modulated pulses, Cambridge, MA: MS Thesis, Electrical Engineering Department, Massachusetts Institute of Technology, 1949.
\bibitem{calude} Calude, C. S., Salomaa, K., \& Roblot, T. K. (2011). Finite-state complexity. Theoretical Computer Science, 412(41), 5668-5677.
\bibitem{schaper} S. Schaper and A.A. Louis, The arrival of the frequent: how bias in genotype-phenotype maps can steer populations to local optima 
\textit{PLoS ONE} 9(2): e86635, 2014.
\bibitem{godelmachines} B.R. Steunebrink, J. Schmidhuber, Towards an Actual G\"odel Machine Implementation. In P. Wang, B. Goertzel, eds., \textit{Theoretical Foundations of Artificial General Intelligence}, Springer, 2012.
\bibitem{solerzenil} Soler-Toscano, F., Zenil, H., Delahaye, J. P. \& Gauvrit, N. (2014). Calculating Kolmogorov complexity from the output frequency distributions of small Turing machines. PloS one, 9(5), e96223.
\bibitem{computabilitypaper} F. Soler-Toscano, H. Zenil, J.-P. Delahaye and N. Gauvrit , Correspondence and Independence of Numerical Evaluations of Algorithmic Information Measures, \textit{Computability}, vol. 2, no. 2, pp 125-140, 2013.
\bibitem{2} J.E. Hopcroft, \& J.D. Ullman, \textit{Formal languages and their relation to automata}, 1969.
\bibitem{kirchner} W. Kirchherr and M. Li and P. Vit\'anyi, The Miraculous Universal Distribution, \textit{Mathematical Intelligencer}, 19, 7--15, 1997.
\bibitem{caevolvingby2dkolmo} B.Y. Peled, V.K. Mishra, A.Y. Carmi, Computing by nowhere increasing complexity, arXiv:1710.01654 [cs.IT]
\bibitem{mondragon} J. Rangel-Mondragon, Recognition and Parsing of Context-Free, \url{http://library.wolfram.com/infocenter/MathSource/3128/} Accessed on Aug 15, 2017.
\bibitem{solerzenilcomplexity} F. Soler-Toscano, H. Zenil, A Computable Measure of Algorithmic Probability by Finite Approximations with an Application to Integer Sequences, \textit{Complexity} (accepted).
\bibitem{schmid1} J. Schmidhuber, Optimal Ordered Problem Solver, \textit{Machine Learning}, 54, 211-254, 2004.. 
\bibitem{schmid2} J. Schmidhuber, V. Zhumatiy, M. Gagliolo, Bias-Optimal Incremental Learning of Control Sequences for Virtual Robots. In Groen, et al. (eds) \textit{Proceedings of the 8th conference on Intelligent Autonomous Systems, IAS-8}, Amsterdam, The Netherlands, pp. 658--665, 2004.
\bibitem{3} F. Soler-Toscano, H. Zenil, J.-P. Delahaye, and N. Gauvrit, Small Turing Machines with Halting State: Enumeration and Running on a Blank Tape. \url{http://demonstrations.wolfram.com/SmallTuringMachinesWithHaltingStateEnumerationAndRunningOnAB/}.
Wolfram Demonstrations Project. Published: January 3, 2013.
\bibitem{solomonoff} 
R.J. Solomonoff. A formal theory of inductive inference: Parts 1 and 2. \textit{Information and Control}, 7:1--22 and 224--254, 1964.
\bibitem{compress} M. Li, and P. Vit\'anyi, \textit{An Introduction to Kolmogorov Complexity and Its Applications}, 3rd ed, Springer, N.Y., 2008.
\bibitem{solo} R.J. Solomonoff, Complexity-Based Induction Systems: Comparisons and Convergence Theorems, \textit{IEEE Trans. on Information Theory}, vol 24, No. 4, pp. 422--432, 1978.
\bibitem{solo2} R.J. Solomonoff, The Application of Algorithmic Probability to Problems in Artificial Intelligence, in L.N. Kanal and J.F. Lemmer (eds.), \textit{Uncertainty in Artificial Intelligence}, pp. 473--491, Elsevier, 1986. 
\bibitem{solo3} Solomonoff, R.J. A System for Incremental Learning Based on Algorithmic Probability, \textit{Proceedings of the Sixth Israeli Conference on Artificial Intelligence, Computer Vision and Pattern Recognition}, Dec. 1989,
pp. 515--527. 
\bibitem{wharton} R.M. Wharton, Grammar enumeration and inference, \textit{Information and Control}, 33(3), 253--272, 1977.
\bibitem{wiering} M. Wiering and J. Schmidhuber. Solving, POMDPs using Levin search and EIRA, \textit{In Proceedings
of the International Conference on Machine Learning (ICML)}, pages 534--542, 1996.
\bibitem{wolfram} S. Wolfram, \textit{A New Kind of Science}, Wolfram Media, Champaign, IL., 2002.
\bibitem{algonature} H. Zenil and J-P. Delahaye, On the Algorithmic Nature of the World, In G. Dodig-Crnkovic and M. Burgin (eds), \textit{Information and Computation,} World Scientific Publishing Company, 2010. 
\bibitem{physicaA} H. Zenil, F. Soler-Toscano, K. Dingle and A. Louis, Correlation of Automorphism Group Size and Topological Properties with Program-size Complexity Evaluations of Graphs and Complex Networks, \textit{Physica A: Statistical Mechanics and its Applications,} vol. 404, pp. 341--358, 2014.
\bibitem{seminars} H. Zenil, N.A. Kiani and J. Tegn\'er, Methods of Information Theory and Algorithmic Complexity for Network Biology, \textit{Seminars in Cell and Developmental Biology,} vol. 51, pp. 32-43, 2016.
\bibitem{kolmo2d} H. Zenil, F. Soler-Toscano, J.-P. Delahaye and N. Gauvrit, Two-Dimensional Kolmogorov Complexity and Validation of the Coding Theorem Method by Compressibility, \textit{PeerJ Computer Science}, 1:e23, 2015.

\end{thebibliography}
\end{document}